\documentclass[twocolumn,pre,superscriptaddress,10pt,aps,floatfix]{revtex4-1}

\usepackage[nice]{nicefrac}
\usepackage{graphicx,color}
\usepackage{amsmath,amssymb,bm,bbm}
\usepackage{amsmath}
\usepackage{braket}
\usepackage{bbold}
\usepackage{float}
\usepackage{placeins}
\usepackage{soul}
\usepackage{enumitem}

\usepackage[plainpages=false,pdfpagelabels,colorlinks=true,linkcolor=red,urlcolor=blue,citecolor=blue,pdftitle={},pdfauthor={},pdfdisplaydoctitle=true,pdfduplex=DuplexFlipLongEdge]{hyperref}

\definecolor{darkred}{rgb}{0.90,0.2,0.2}
\definecolor{darkgreen}{rgb}{0,0.60,.2}
\definecolor{darkblue}{rgb}{0.1,0.3,1}
\definecolor{grey}{cmyk}{0,0,0,0.25}
\definecolor{orange}{cmyk}{0,0.6,0.8,0}

\begin{document}

\title{Generalized Thermalization in Quantum-Chaotic Quadratic Hamiltonians}

\author{Patrycja  \L yd\.{z}ba}
\affiliation{Institute of Theoretical Physics, Wroclaw University of Science and Technology, 50-370 Wroc{\l}aw, Poland}
\author{Marcin Mierzejewski}
\affiliation{Institute of Theoretical Physics, Wroclaw University of Science and Technology, 50-370 Wroc{\l}aw, Poland}
\author{Marcos Rigol}
\affiliation{Department of Physics, The Pennsylvania State University, University Park, Pennsylvania 16802, USA}
\author{Lev Vidmar}
\affiliation{Department of Theoretical Physics, J. Stefan Institute, SI-1000 Ljubljana, Slovenia}
\affiliation{Department of Physics, Faculty of Mathematics and Physics, University of Ljubljana, SI-1000 Ljubljana, Slovenia}

\begin{abstract}
Thermalization (generalized thermalization) in nonintegrable (integrable) quantum systems requires two ingredients: equilibration and agreement with the predictions of the Gibbs (generalized Gibbs) ensemble. We prove that observables that exhibit eigenstate thermalization in single-particle sector equilibrate in many-body sectors of quantum-chaotic quadratic models. Remarkably, the same observables do not exhibit eigenstate thermalization in many-body sectors (we establish that there are exponentially many outliers). Hence, the generalized Gibbs ensemble is generally needed to describe their expectation values after equilibration, and it is characterized by Lagrange multipliers that are smooth functions of single-particle energies.
\end{abstract}
\maketitle

\textit{Introduction.}---Over the past 15 years, we have improved significantly our understanding of quantum dynamics in isolated many-body quantum systems~\cite{dalessio_kafri_16, Eisert2015, mori_ikeda_18, deutsch_18}. A paradigmatic setup for these studies is the quantum quench, in which a sudden change of a tuning parameter pushes the system far from equilibrium. Following quantum quenches, observables in nonintegrable systems have been found to equilibrate to the predictions of the Gibbs ensemble (GE)~\cite{rigol_dunjko_08, dalessio_kafri_16}, while in integrable systems they have been found to equilibrate to the predictions of the generalized Gibbs ensemble (GGE)~\cite{rigol_dunjko_07, vidmar16}. The validity of the GGE has been tested in many theoretical studies of integrable models that are mappable onto quadratic ones~\cite{rigol_dunjko_07, cazalilla_06, rigol_muramatsu_06, iucci_cazalilla_09, Cassidy_2011, calabrese_essler_11, Gramsch_2012, calabrese_essler_12b, essler_evangelisti_12, caux_essler_13, vidmar16} and integrable models that are not mappable onto quadratic ones~\cite{caux_konik_12, kormos_shashi_13, pozsgay_13, fagotti_collura_14, nardis_wouters_14, wouters_denardis_14, pozsgay_mestyan14, Mierzejewski_2015, ilievski_medenjak_15, ilievski_denardis_15, piroli_vernier_17, fagotti_maric_22} (see Ref.~\cite{calabrese_essler_review_16} for reviews). The GGE is also a starting point for the recently introduced~\cite{bertini2016transport, castro2016emergent} and experimentally tested~\cite{schemmer2019generalized, malvania_zhang_21} theory of generalized hydrodynamics.

Quadratic fermionic models, which are central to understanding a wide range of phenomena in condensed matter physics, can be thought as being a special (noninteracting) class of integrable models. Their Hamiltonians consist of bilinear forms of creation and annihilation operators. The infinite-time averages of one-body observables after quenches in these models are always described by GGEs~\cite{vidmar16, Ziraldo_2012, Ziraldo_2013, He_2013}. However, there are one-body observables that generically fail to equilibrate because the one-body density matrix evolves unitarily~\cite{wright_rigol_14}, i.e., generalized thermalization fails to occur. Such equilibration failures have been discussed in the context of localization in real~\cite{Gramsch_2012, Ziraldo_2012, Ziraldo_2013, He_2013} and momentum~\cite{He_2013, wright_rigol_14} space. Equilibration in quadratic models has been argued to occur for local observables in the absence of real-space localization. In particular, it has been shown to occur for initial states that are ground states of local Hamiltonians~\cite{Cramer_2008}, as well as for initial states that exhibit sufficiently rapidly decaying correlations in real space~\cite{Gluza_2016, murthy19, Gluza_2019}. 

In this Letter, we show that there is a broad class of quadratic fermionic models for which generalized thermalization is ensured by the properties of the Hamiltonian. Hence, it is robust and resembles thermalization (generalized thermalization), which occurs in interacting nonintegrable (interacting integrable) models. The class in question is that of quantum-chaotic quadratic (QCQ) Hamiltonians, namely, quadratic Hamiltonians that exhibit single-particle quantum chaos~\cite{lydzba_rigol_21, lydzba_zhang_21}. Paradigmatic examples of local QCQ models are the three-dimensional (3D) Anderson model in the delocalized regime~\cite{lydzba_rigol_21, suntajs_prosen_21} and chaotic tight-binding billiards~\cite{ulcakar_vidmar_22}, while their nonlocal counterparts include variants of the quadratic Sachdev-Ye-Kitaev (SYK2) model~\cite{lydzba_rigol_20, liu_chen_18} and the power-law random banded matrix model in the delocalized regime~\cite{lydzba_rigol_20}. The single-particle sector of those models exhibits random-matrix-like statistics of the energy levels~\cite{altshuler_shklovskii_86, altshuler_zharekeshev_88, suntajs_prosen_21}, as well as single-particle eigenstate thermalization~\cite{lydzba_zhang_21}, i.e., the matrix elements of properly normalized one-body observables $\hat {\cal O}$~\footnote{Namely, for lattice systems such as the ones of interest here, satisfying $\frac{1}{V}{\rm Tr}\{ \hat {\cal O}^2 \}=1$, where $V$ is the number of lattice sites.} in the {\it single-particle} energy eigenkets are described by the eigenstate thermalization hypothesis ansatz~\cite{srednicki_99, dalessio_kafri_16}
\begin{equation} \label{def_eth_ansatz}
\langle \alpha |\hat {\cal O} |\beta\rangle = {\cal O}(\bar \epsilon) \delta_{\alpha\beta} + \rho(\bar \epsilon)^{-1/2} {\cal F}_{\cal O}(\bar \epsilon, \omega) R^{\cal O}_{\alpha\beta}\;,
\end{equation}
where $\bar \epsilon = (\epsilon_\alpha + \epsilon_\beta)/2$, $\omega = \epsilon_\beta-\epsilon_\alpha$, ${\cal O}(\bar \epsilon)$ and ${\cal F}_{\cal O}(\bar \epsilon, \omega)$ are smooth functions of their arguments, while $\rho(\bar \epsilon) = \delta N/\delta \epsilon|_{\bar \epsilon}$ is the single-particle density of states (typically proportional to the volume) at energy $\bar \epsilon$. The distribution of matrix elements is described by the random variable $R^{\cal O}_{\alpha\beta}$, which has zero mean and unit variance. The {\it many-body} energy eigenstates, on the other hand, exhibit eigenstate entanglement properties typical of Gaussian states~\cite{lydzba_rigol_20, lydzba_rigol_21, bianchi_hackal_21, bianchi_hackl_22, liu_chen_18}; see also \cite{Projected_1}. 

We prove that single-particle eigenstate thermalization ensures equilibration in many-body sectors of QCQ Hamiltonians, and we also prove that eigenstate thermalization does not occur in those sectors. We then show that the GGE is needed to describe observables after equilibration, and that it is characterized by the Lagrange multipliers that are smooth functions of the single-particle energies. The latter is also a consequence of single-particle eigenstate thermalization. Our analytical results are tested numerically in QCQ Hamiltonians and contrasted with results obtained for quadratic models that are not quantum chaotic.

\textit{Quantum quench and equilibration.}---We consider a quantum quench setup; the system is prepared in an initial many-body pure state $|\Psi_0\rangle$ and evolves unitarily under a quadratic Hamiltonian $\hat{H}=\sum_{i,j=1}^V h_{ij}\hat{c}^\dagger_i\hat{c}^{}_j$, where $\hat{c}^\dagger_i$ ($\hat{c}^{}_i$) creates (annihilates) a spinless fermion at site $i$, and $V$ denotes the number of lattice sites. In what follows, we use uppercase (lowercase) Greek letters to denote quantum states in the many-body (single-particle) Hilbert space. One can diagonalize $\hat{H}$ via a unitary transformation of the creation and annihilation operators, $\hat{H}=\sum_{\alpha} \epsilon_{\alpha} \hat{f}_\alpha^\dagger \hat{f}^{}_\alpha$. The single-particle energy eigenstates, with eigenenergies~$\epsilon_\alpha$, can be written as $|\alpha\rangle \equiv \hat f_\alpha^\dagger |\emptyset\rangle$. The many-body energy eigenstates, with eigenenergies $E_\Omega=\sum_{\left\{\alpha\right\}} \epsilon_{\alpha}$, can be written as $|\Omega\rangle=\prod_{\left\{\alpha\right\}}\hat{f}^\dagger_{\alpha}|\emptyset\rangle$, where $\left\{\alpha\right\}$ is the set of $N$ occupied $|\alpha\rangle$ for any given lattice filling $\bar{n}=N/V$. Any initial many-body pure state can be written as $|\Psi_0\rangle=\sum_\Omega \langle\Omega|\Psi_0\rangle |\Omega\rangle$.

Our focus is on one-body observables with rank ${\bf O}(1)$~\footnote{Namely, observables that have an ${\bf O}(1)$ number of nondegenerate eigenvalues in the single-particle spectrum.}, such as site and quasimomentum occupations, which are experimentally relevant and have the following form $\hat{O}=\sum_{\alpha\beta} O_{\alpha\beta} \hat{f}_\alpha^\dagger \hat{f}^{}_\beta$ with $O_{\alpha\beta} = \langle \alpha| \hat{O} |\beta\rangle$. Their time evolution can be written in the many-body basis as $\langle\hat{O}(t)\rangle = \sum_\Omega \langle\Omega|e^{-i\hat{H}t} \hat{\rho}_0 e^{i\hat{H}t} \hat{O}| \Omega\rangle$, where $\hat{\rho}_0 = |\Psi_0\rangle \langle\Psi_0|$ is the density matrix of the initial state. In quadratic models, we can write it using the single-particle basis as
\begin{equation} \label{def_Ot}
    \langle\hat{O}(t)\rangle  = \sum_{\alpha}\langle\alpha|e^{-i\hat{H}t} \hat R e^{i\hat{H}t}\hat{O}|\alpha\rangle\\
     = \sum_{\alpha,\beta=1}^V R_{\alpha\beta} O_{\beta\alpha} e^{i\omega_{\beta\alpha}t},
\end{equation}
where $\hat{R} = \sum_{\alpha\beta} R_{\alpha\beta} \hat f_\alpha^\dagger \hat f^{}_\beta$ is the one-body density matrix of the initial state, with $R_{\alpha\beta} = \langle\Psi_0|\hat{f}_{\beta}^\dagger\hat{f}^{}_{\alpha}|\Psi_0\rangle$~\cite{Venuti_13, Venuti_2015}, and $\omega_{\beta\alpha}=\epsilon_\beta - \epsilon_\alpha$.

The infinite time average of $\langle\hat{O}(t)\rangle$, for a nondegenerated single-particle spectrum, is given by
\begin{equation}
\label{def_inf}
    \overline{\langle\hat{O}(t)\rangle} \equiv \lim_{\tau\to\infty} \frac{1}{\tau} \int_{0}^{\tau} \langle\hat{O}(t)\rangle dt=
    \sum_{\alpha} O_{\alpha\alpha} R_{\alpha\alpha}.
\end{equation}
The density matrix in the GGE is defined as $\hat{\rho}_\text{GGE}=\frac{1}{Z_\text{GGE}} e^{-\sum_\alpha \lambda_\alpha \hat{I}_\alpha}$ with $Z_\text{GGE}=\text{Tr}[ e^{-\sum_\alpha \lambda_\alpha \hat{I}_\alpha}]$, the constants of motion being $\hat{I}_\alpha=\hat{f}_\alpha^\dagger \hat{f}^{}_\alpha$, and the Lagrange multipliers fixed such that 
$R_{\alpha \alpha}=\text{Tr}[\hat{\rho}_\text{GGE} \hat I_\alpha]$.
Therefore, the infinite-time average of $\langle\hat{O}(t)\rangle$ is reproduced by the GGE prediction~\cite{vidmar16, Ziraldo_2013, He_2013}
\begin{equation}
    \overline{\langle\hat{O}(t)\rangle}=
     \sum_{\alpha} O_{\alpha\alpha} \text{Tr}[\hat{\rho}_\text{GGE}\hat{I}_{\alpha}]=
     \text{Tr}[\hat{\rho}_\text{GGE}\hat{O}],
\end{equation}
where we have used that $\hat{\rho}_\text{GGE}$ is diagonal in the single-particle energy eigenbasis, so that $\text{Tr}[\hat{\rho}_\text{GGE}\sum_{\alpha\beta} O_{\alpha\beta}\hat{f}_\alpha^\dagger \hat{f}^{}_\beta]=\text{Tr}[\hat{\rho}_\text{GGE}\sum_{\alpha} O_{\alpha\alpha}\hat{f}_\alpha^\dagger \hat{f}^{}_\alpha]$.

Given that the infinite-time averages are guaranteed to be described by the GGE, all one needs for generalized thermalization to occur is the temporal fluctuations about the infinite-time average to vanish in the thermodynamic limit. The temporal fluctuations can be characterized by the variance~\cite{dalessio_kafri_16}
\begin{equation} \label{def_sigmat}
\sigma_t^2 = \overline{\langle\hat{O}(t)\rangle^2} - \overline{\langle\hat{O}(t)\rangle}^2 \;.
\end{equation}
Recall that the standard derivation of the upper bound for $\sigma_t^2$, which is based on the time evolution written in the many-body basis, requires that there are no gap degeneracies in the many-body spectrum~\cite{dalessio_kafri_16}. This condition need not be fulfilled in quadratic models. The derivation that we provide below, which is based on Eq.~\eqref{def_Ot}, requires the absence of gap degeneracies in the single-particle spectrum. The latter is satisfied by QCQ Hamiltonians. Specifically, one can write
\begin{equation}
    \overline{\langle\hat{O}(t)\rangle^2}= 
    \sum_{\alpha,\beta,\omega,\rho} O_{\beta\alpha}O_{\rho\omega}
    R_{\alpha\beta}R_{\omega\rho}
    \overline{e^{i\left(\epsilon_\beta-\epsilon_\alpha+\epsilon_\rho-\epsilon_\omega\right)t}}\;,
\end{equation}
which simplifies to (see Ref.~\cite{suppmat})
\begin{equation}
    \overline{\langle\hat{O}(t)\rangle^2}=\sum_{\alpha\neq\beta} |O_{\alpha\beta}|^2 |R_{\alpha\beta}|^2+\overline{\langle\hat{O}(t)\rangle}^2\;.
\end{equation}
We can therefore define the upper bound for the variance
\begin{equation} \label{eq_bound1}
    \sigma_t^2 =\sum_{\alpha\ne\beta=1}^V |O_{\alpha\beta}|^2 |R_{\alpha\beta}|^2 \le \text{max}\left\{|O_{\alpha\beta}|^2\right\}\sum_{\alpha=1}^V (R^2)_{\alpha\alpha}\,.
\end{equation}
Since the eigenvalues of $\hat{R}$ belong to the interval $[0,1]$, one can replace $R^2 \to R$ in Eq.~(\ref{eq_bound1}), and we obtain
\begin{equation} \label{eq_bound2}
    \sigma_t^2 \leq \text{max}\left\{V |O_{\alpha\beta}|^2\right\} \frac{1}{V}\sum_{\alpha=1}^V R_{\alpha\alpha} = \text{max}\left\{V |O_{\alpha\beta}|^2\right\} \bar{n} \;,
\end{equation}
where we used that $\sum_{\alpha} R_{\alpha\alpha} = \langle\psi_0| \sum_\alpha \hat f_\alpha^\dagger \hat f_\alpha |\psi_0\rangle = N$. Because the properly normalized one-body observables with rank ${\bf O}(1)$~\cite{Note1, Note2} can be written as $\hat {\cal O}\simeq \hat O \sqrt{V}$~\cite{lydzba_zhang_21}, single-particle eigenstate thermalization in QCQ models results in $\text{max}\left\{V |O_{\alpha\beta}|^2\right\} \propto 1/V$. Hence, the equilibration of these one-body observables is guaranteed in the thermodynamic limit. Notice that the polynomial scaling of the upper bound for $\sigma_t^2$ with the system size is independent of the details of the quantum quench, like the energy of the initial state $|\Psi_0\rangle$ or the filling factor $\bar{n}$. 
The above analysis can be extended to one-body operators that have rank ${\bf O}(V)$.
Furthermore, in Ref.~\cite{suppmat} we show that equilibration also occurs for $q$-body observables ($q=2,3,...$) that are products of one-body observables, all of which exhibit single-particle eigenstate thermalization. Remarkably, our analysis applies to arbitrary initial states~\cite{suppmat}.

\textit{Numerical tests of equilibration.}---We consider local Hamiltonians that can be written as
\begin{equation} \label{def_H1}
\hat{H}_1=-\sum_{\langle i,j\rangle}\hat{c}^\dagger_i\hat{c}_j + \sum_{i=1}^{V} \varepsilon_i\hat{c}^\dagger_i\hat{c}_i  \;.
\end{equation}
The first term describes hoppings between nearest neighbor sites, and $\varepsilon_i$ is the on-site potential. We focus on the 3D Anderson model on a cubic lattice with periodic boundary conditions, for which $\varepsilon_i = (W/2) r_i$ with $r_i$ being a random number drawn from a uniform distribution in the interval $\left[-1,1\right]$~\cite{Anderson_58}. We study dynamics in the two regimes of this model (which has a transition at $W_c \approx 16.5$~\cite{Slevin_2014,slevin_ohtsuki_18}), at the $W=5$ (delocalized, QCQ~\cite{lydzba_rigol_21}), and $W=25$ (localized) points. For the preparation of initial states in quantum quenches, which are always taken to be ground states in this work, we introduce a 3D superlattice model with $\varepsilon_i = \pm \mathcal{W}$ in Eq.~(\ref{def_H1}), where the sign alternates between nearest neighbor sites. This 3D superlattice model allows us to create highly nonthermal distributions of momenta in the initial state (in the spirit of the quantum Newton's cradle experiment~\cite{Kinoshita2006}). We complement our analysis with a quadratic model that is not quantum chaotic, i.e., 1D noninteracting fermions in a homogeneous lattice with open boundary conditions [$\varepsilon_i=0$ in Eq.~(\ref{def_H1})]. To prepare the initial states for the quenches, we use the Aubry-Andr\'e model [$\varepsilon_i=-\Lambda \cos(2\pi \sigma i)$ with $\sigma = (\sqrt{5}-1)/2$] in Eq.~(\ref{def_H1}).

We also consider a paradigmatic nonlocal QCQ model, the SYK2 model in the Dirac fermion formulation~\cite{Sachdev_93},
\begin{equation} \label{def_H2}
\hat{H}_2=\sum_{i,j=1}^{V}\left[(1-\gamma)a_{ij}+\gamma b_{ij}\right]\hat{c}^\dagger_i\hat{c}_j,
\end{equation}
where the diagonal (off-diagonal) elements of the matrices ${\bf a}$ and ${\bf b}$ are real normally distributed random numbers with zero mean and $2/V$ ($1/V$) variance, while $\gamma\in\left[0,1\right]$. The choice of an unconventional form of the SYK2 Hamiltonian (as a sum of two one-body operators) allows us to distinguish between weak and strong quantum quenches, as explained in Ref.~\cite{suppmat}.

\begin{figure}[!t]
\includegraphics[width=0.98\columnwidth]{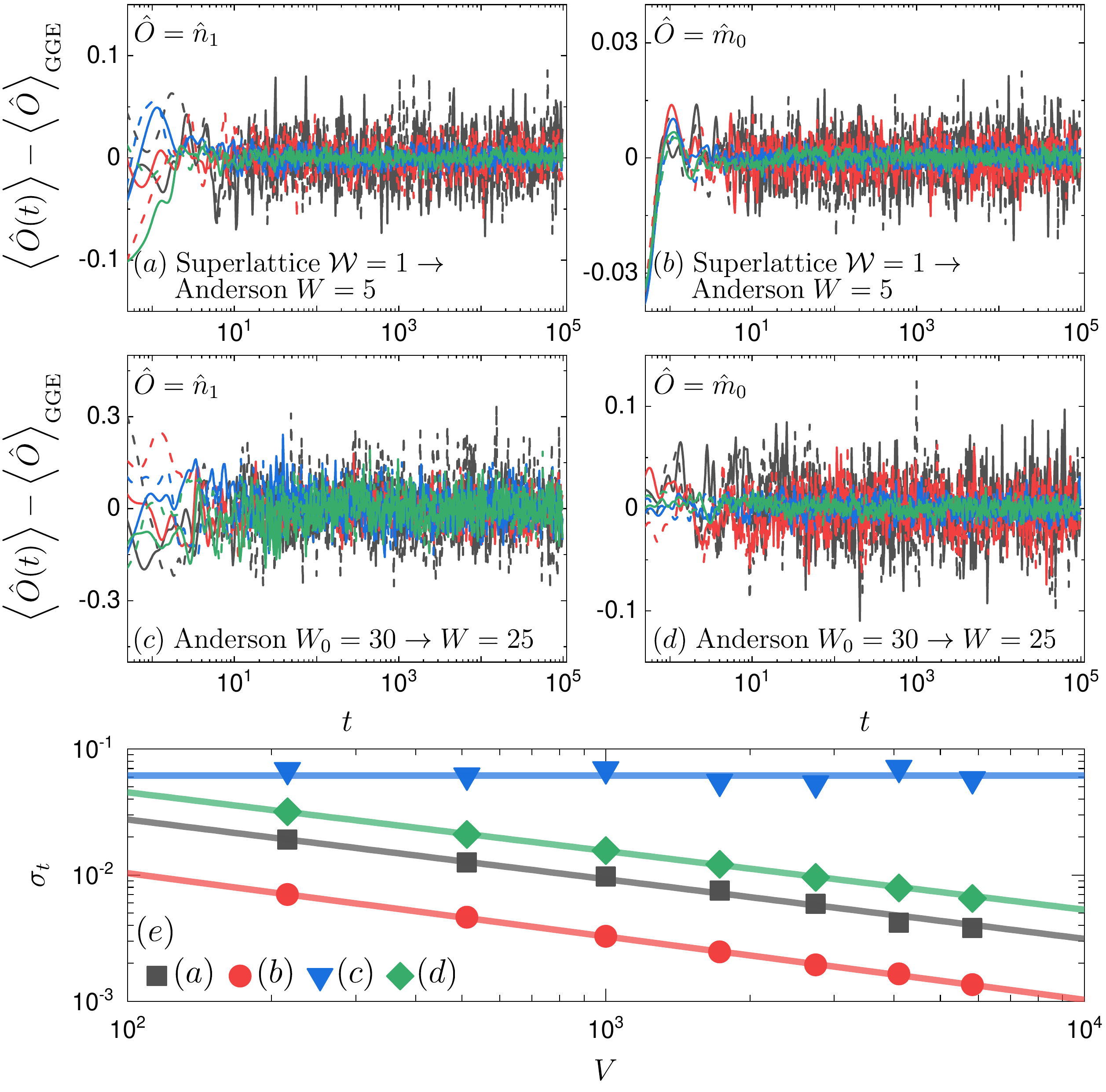}
\vspace{-0.2cm}
\caption{(a)--(d) Time evolution of $\langle\hat{O}(t)\rangle - \langle\hat{O}\rangle_\text{GGE}$ after quantum quenches in 3D models. The numerical results for system with $V=6^3,\, 8^3,\, 14^3,$ and $18^3$ are marked with black, red, blue, and green, respectively. We show results for two (solid and dashed) quench realizations for each $V$. (a), (b) Quenches from the 3D superlattice model at ${\cal W} = 1$ and $\bar{n} = 1/4$ to the 3D Anderson model at $W=5$. (c), (d) Quenches from the 3D Anderson model at $W_0 = 30$ and $\bar{n} = 1/2$ to the same model (with a different disorder realization) at $W=25$. Two operators are considered (a), (c) $\hat n_1$ and (b), (d) $\hat m_0$. (e) Temporal fluctuations $\sigma_t$ calculated within the time interval $t\in[10^2, 10^5]$ and averaged over $20$ quench realizations. The lines show the outcome of two parameter fits $\kappa/V^\zeta$. We get $\zeta\in[0.46,0.5]$ for (a), (b), and (d).}
\label{fig1}
\end{figure}

In Fig.~\ref{fig1}, we show results of numerical tests of equilibration for two observables, the occupation of a lattice site, $\hat{n}_{1} = \hat{c}_{1}^\dagger \hat{c}^{}_{1}$, and the occupation of the zero quasimomentum mode, $\hat{m}_{0}=\frac{1}{V}\sum_{ij}\hat{c}_{i}^\dagger\hat{c}_{j}$. Specifically, we plot the time evolution of $\langle\hat{O}(t)\rangle-\langle\hat{O}\rangle_\text{GGE}$ in Figs.~\ref{fig1}(a)--\ref{fig1}(d), while the temporal fluctuations $\sigma_t$ as functions of $V$ are shown in Fig.~\ref{fig1}(e). For the quench from the 3D superlattice model at $\mathcal{W}=1$ to the 3D Anderson model at $W=5$, see Figs.~\ref{fig1}(a) and~\ref{fig1}(b), the temporal fluctuations $\sigma_t$ of both observables decrease with increasing system size, and a scaling $\sigma_t \propto V^{-\zeta}$ with $\zeta \approx 0.5$ is observed in Fig.~\ref{fig1}(e). An exponent $\zeta = 0.5$ is expected because $V |O_{\alpha\beta}|^2 \propto 1/V$ for those observables~\cite{lydzba_zhang_21}. In contrast, for the quench from the 3D Anderson model at $W_0=30$ to the same model (with a different disorder realization) at $W=25$ in Fig.~\ref{fig1}(c) [Fig.~\ref{fig1}(d)], the temporal fluctuations $\sigma_t$ do not decrease (do decrease) with increasing system size for $\hat n_1$ ($\hat m_0$), and a scaling $\sigma_t \propto V^{-\zeta}$ with $\zeta \approx 0$ ($\zeta \approx 0.5$) is observed in Fig.~\ref{fig1}(e). This a consequence of the fact that $\hat{m}_{0}$, but not $\hat{n}_{1}$, exhibits signatures of single-particle eigenstate thermalization in the localized regime of the 3D Anderson model~\cite{lydzba_zhang_21}.
The results for $\hat n_1$ show that equilibration is not guaranteed for quadratic Hamiltonians that are not quantum chaotic.
Qualitatively similar results to those for $W=25$ in the 3D Anderson model were reported in the presence of real-space localization in the 1D Anderson model~\cite{Ziraldo_2012, Ziraldo_2013}, and in the 1D Aubry-Andr\'e model~\cite{Gramsch_2012, He_2013}.

\textit{Stationary state.}---Since eigenstate thermalization occurs in single-particle eigenstates of QCQ models, it is natural to wonder whether it also occurs in the many-body eigenstates of those models. If this is the case, the predictions of the GGE will be identical to those of the GE in the thermodynamic limit, $\langle\hat{O}\rangle_\text{GGE} = \langle\hat{O}\rangle_\text{GE}$, where $\langle\hat{O}\rangle_\text{GGE} = \text{Tr}[\hat{\rho}_\text{GGE}\hat{O}]$ and $\langle\hat{O}\rangle_\text{GE} = \text{Tr}[\hat{\rho}_\text{GE}\hat{O}]$, with $\hat{\rho}_\text{GE} = \frac{1}{Z_\text{GE}} e^{-\sum_\alpha (\epsilon_\alpha-\mu)/(k_BT)\, \hat{f}^\dagger_\alpha \hat{f}^{}_{\alpha}}$, and $Z_{\rm GE} = {\rm Tr}[e^{-\sum_\alpha (\epsilon_\alpha-\mu)/(k_B T) \hat{f}^\dagger_\alpha \hat{f}_{\alpha}}]$. $k_B$, $T$, and $\mu$ are the Boltzmann constant, the temperature, and the chemical potential, respectively.

\begin{figure}[!t]
\includegraphics[width=0.98\columnwidth]{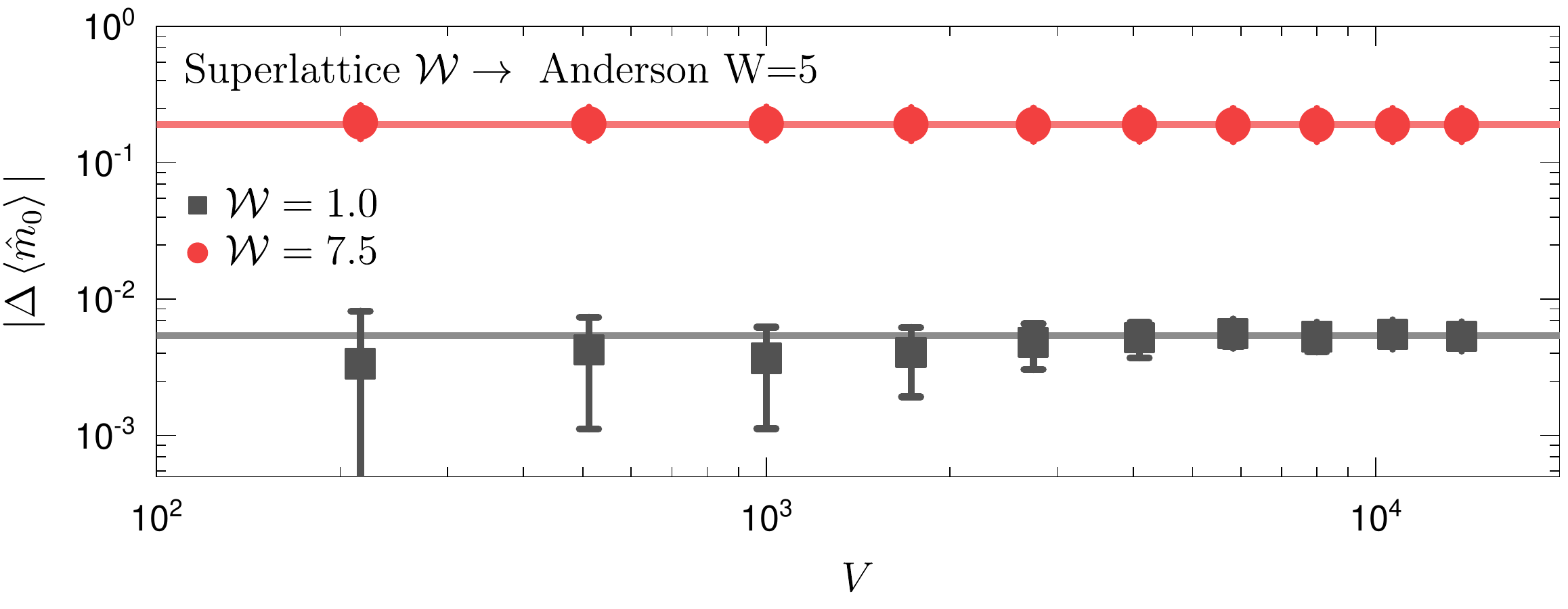}
\vspace{-0.3cm}
\caption{Finite-size scaling of the difference $|\Delta\langle\hat{m}_{0}\rangle|$ for quenches from the 3D superlattice model with $\mathcal{W}\in\left\{1,7.5\right\}$ and $\bar{n} = 1/4$ to the 3D Anderson model with $W = 5$. Horizontal lines mark the mean values for the five largest system sizes $V \in \left\{16^3, \, ..., \, 24^3 \right\}$. The results were averaged over $M=100$ quench realizations. The error bars are standard deviations $\sigma = \left(\sum_{i=1}^{M} |\Delta\langle \hat{m}_{0}\rangle|_i^{2}/{M} - (\sum_{i=1}^{M}| \Delta\langle \hat{m}_{0}\rangle|_i/{M})^2\right)^{1/2}$.}
\label{fig2}
\end{figure}

We address this question in the context of the quenches to the 3D Anderson model with $W=5$. We focus on $\hat{m}_{0}$. The finite-size scaling of the difference $|\Delta\langle\hat{m}_{0}\rangle|= |\langle\hat{m}_0\rangle_\text{GGE}-\langle\hat{m}_0\rangle_\text{GE}|$ is reported in Fig.~\ref{fig2}. Each point was calculated for a single quench realization, and then averaged over $100$ quench realizations. It is apparent that the difference $|\Delta \langle\hat{m}_{0}\rangle|$ rapidly converges to a nonzero value. Therefore, the GGE is expected to be different from the GE in the thermodynamic limit. (Qualitatively similar results as in this section and the previous one were obtained for other models, quenches and observables.)

\textit{Absence of eigenstate thermalization in many-body energy eigenstates.}---The numerical results from the previous section suggest that the many-body eigenstates of QCQ Hamiltonians do not exhibit eigenstate thermalization [the infinite time averages from Eq.~(\ref{def_inf}) disagree with the predictions of the GE]. We can understand this analytically as follows (see Ref.~\cite{suppmat} for a proof).

The diagonal matrix elements of $\hat{O}$ in the many-body energy eigenstates $|\Omega\rangle$ can be written as
\begin{equation}
\label{eq_diag_mb_0}
    \langle\Omega|\hat{O}|\Omega\rangle = \sum_{\alpha,\beta=1}^{V} O_{\alpha\beta}\langle\Omega|\hat{f}^\dagger_{\alpha}\hat{f}^{}_{\beta}|\Omega\rangle=\sum_{\alpha=1}^{V} O_{\alpha\alpha}\langle\Omega|\hat{f}^\dagger_{\alpha}\hat{f}^{}_{\alpha}|\Omega\rangle\;,
\end{equation}
where the expectation values $\langle\Omega| \hat{f}^\dagger_{\alpha} \hat{f}^{}_{\alpha} |\Omega\rangle$ are equal 0 or 1. Hence, the behavior of the diagonal many-body matrix elements [and so the infinite-time averages from Eq.~(\ref{def_inf})] is governed by an extensive (in $V$) sum of the diagonal single-particle matrix elements $O_{\alpha\alpha}$.

The diagonal matrix elements $O_{\alpha\alpha}$ exhibit ${\bf O}(1/V)$ fluctuations about their smooth function $\mathcal{O}(\epsilon_{\alpha})$~\cite{lydzba_zhang_21}. For simplicity, let us consider $\mathcal{O}(\epsilon_{\alpha})=0$. We can build many-body eigenstates~$|\Omega\rangle$, for which $V/a$ and $V/b$ of $ O_{\alpha\alpha} \langle\Omega| \hat{f}^\dagger_{\alpha} \hat{f}^{}_{\alpha} |\Omega\rangle$ are positive and negative, respectively. The corresponding diagonal matrix elements read
\begin{equation}
\langle\Omega|\hat{O}|\Omega\rangle=\sum_{\beta=1}^{V/a}|O_{\beta\beta}|-\sum_{\beta=1}^{V/b}|O_{\beta\beta}| \sim \left(\frac{V}{a}-\frac{V}{b}\right){\bf O}\left(\frac{1}{V}\right),
\end{equation}
where $1/a+1/b=\bar{n}$. These diagonal matrix elements are ${\bf O}(1)$ when $a\neq b$, so they do not approach the microcanonical average in the thermodynamic limit. Furthermore, the number of such many-body states increases exponentially with the system size
\begin{equation}
\begin{split}
    \mathcal{N} & =
    \begin{pmatrix}
    V/2\\ 
    V/a
    \end{pmatrix}
    \begin{pmatrix}
    V/2\\ 
    V/b
    \end{pmatrix}
    \ge
   \left[\frac{V/2}{V/a}\right]^\frac{V}{a} \left[\frac{V/2}{V/b}\right]^\frac{V}{b}
    =2^{\kappa V}\;,
\end{split}
\end{equation}
where we have introduced $\kappa =\frac{1}{a}\log_2\left(\frac{a}{2}\right)+\frac{1}{b}\log_2\left(\frac{b}{2}\right)$.

\textit{Smoothness of Lagrange multipliers.}---To conclude, let us explore the properties of the GGE in QCQ Hamiltonians. Note that whenever the Lagrange multipliers are linear functions of the single-particle energies, i.e., $\lambda_\alpha=(\epsilon_\alpha-\mu)/(k_B T)$, the GGE is the same as the GE.

\begin{figure}[!t]
\includegraphics[width=0.98\columnwidth]{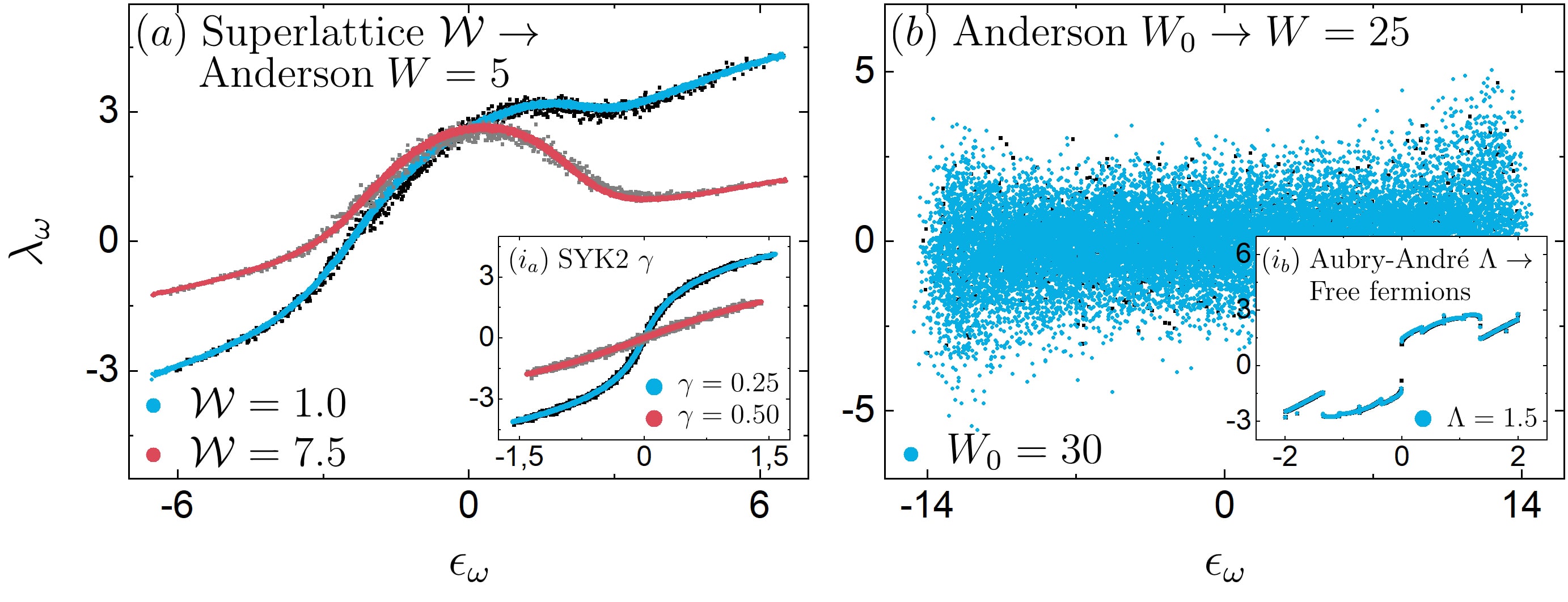}
\vspace{-0.2cm}
\caption{Lagrange multipliers $\lambda_\omega$ plotted versus single-particle energies $\epsilon_\omega$. Black and gray (blue and red) points depict results for $V=10^3$ ($V=28^3$) and a single quench realization. These quenches are (a) the 3D superlattice model with ${\cal W}=1$ and 7.5 to the 3D Anderson model with $W=5$ (main panel), and the change of ${\bf b}$ to a new random realization in the SYK2 model with $\gamma=0.25$ and 0.5 (inset); (b) the 3D Anderson model at $W_0=30$ to the same model (with a different disorder realization) at $W=25$ (main panel), and the Aubry-Andr\'e model with $\Lambda=1.5$ to free fermions (inset). In all cases $\bar{n}=1/2$, except for the main panel in (a) where $\bar{n}=1/4$.}
\label{fig3}
\end{figure}

The Lagrange multipliers $\lambda_\alpha$ are plotted as functions of the single-particle energies $\epsilon_\alpha$ in Fig.~\ref{fig3}(a) for quenches in which the final Hamiltonian exhibits single-particle quantum chaos: the 3D Anderson model with $W=5$ (main panel) and the SYK2 model (inset). It is notable that $\lambda_\alpha$ are smooth functions of $\epsilon_\alpha$, and that they are not linear in $\epsilon_\alpha$, even in the quench within the SYK2 model with an arbitrary $\gamma<1$ (the exception is $\gamma=1$, which is at ``infinite temperature'', see Ref.~\cite{suppmat}). In Fig.~\ref{fig3}(b), we plot the Lagrange multipliers $\lambda_\alpha$ vs $\epsilon_\alpha$ for quenches in which the final Hamiltonian does not exhibit single-particle quantum chaos: the 3D Anderson model in the localized regime with $W=25$ (main panel) and 1D noninteracting fermions in a homogeneous potential (inset). In the former, $\lambda_\alpha$ exhibits fluctuations that do not appear to vanish when increasing system size $V$, while in the latter $\lambda_\alpha$ exhibits jumps. In Ref.~\cite{suppmat}, we quantify the eigenstate-to-eigenstate fluctuations $\delta\lambda_\alpha = \lambda_\alpha - \lambda_{\alpha-1}$, and show numerically and analytically that $\lambda_\alpha$ is a smooth function of $\epsilon_\alpha$ for QCQ Hamiltonians.

\textit{Summary.}---Generalized thermalization is expected to occur for interacting integrable models. Here, we proved that it is guaranteed to occur for quadratic Hamiltonians that exhibit single-particle eigenstate thermalization, namely, for QCQ Hamiltonians. Furthermore, we showed that the many-body eigenstates of QCQ Hamiltonians do not exhibit eigenstate thermalization. Consequently, the GGE is generally needed to describe the expectation values of observables after equilibration, and we showed that it is characterized by Lagrange multipliers that are smooth functions of the single-particle energies.

\acknowledgements 
We acknowledge the support of the National Science Centre, Poland via project 2020/37/B/ST3/00020 (M.M.), the National Science Foundation, Grant No.~2012145 (M.R.), and the Slovenian Research Agency (ARRS), Research Core Fundings Grants P1-0044, J1-1696 and N1-0273 (L.V.).

\FloatBarrier
\bibliographystyle{biblev1}
\bibliography{references}

\begin{thebibliography}{10}
\expandafter\ifx\csname url\endcsname\relax
  \def\url#1{{\tt #1}}\fi
\expandafter\ifx\csname urlprefix\endcsname\relax\def\urlprefix{URL }\fi
\expandafter\ifx\csname bibinfo\endcsname\relax\def\bibinfo#1#2{#2}\fi
\expandafter\ifx\csname eprint\endcsname\relax\def\eprint#1{\url{#1}}\fi

\bibitem{dalessio_kafri_16}
\bibinfo{author}{L.~D'Alessio}, \bibinfo{author}{Y.~Kafri},
  \bibinfo{author}{A.~Polkovnikov}, and \bibinfo{author}{M.~Rigol},
  \bibinfo{title}{From quantum chaos and eigenstate thermalization to
  statistical mechanics and thermodynamics},
  \bibinfo{journal}{\href{http://dx.doi.org/10.1080/00018732.2016.1198134}{Adv.
  Phys.}} \href{http://dx.doi.org/10.1080/00018732.2016.1198134}{{\bf
  \bibinfo{volume}{65}}, \bibinfo{pages}{239}}
  (\href{http://dx.doi.org/10.1080/00018732.2016.1198134}{\bibinfo{year}{2016}}).

\bibitem{Eisert2015}
\bibinfo{author}{J.~Eisert}, \bibinfo{author}{M.~Friesdorf}, and
  \bibinfo{author}{C.~Gogolin}, \bibinfo{title}{Quantum many-body systems out
  of equilibrium},
  \bibinfo{journal}{\href{http://dx.doi.org/10.1038/nphys3215}{Nat. Phys.}}
  \href{http://dx.doi.org/10.1038/nphys3215}{{\bf \bibinfo{volume}{11}},
  \bibinfo{pages}{124}}
  (\href{http://dx.doi.org/10.1038/nphys3215}{\bibinfo{year}{2015}}).

\bibitem{mori_ikeda_18}
\bibinfo{author}{T.~Mori}, \bibinfo{author}{T.~N. Ikeda},
  \bibinfo{author}{E.~Kaminishi}, and \bibinfo{author}{M.~Ueda},
  \bibinfo{title}{Thermalization and prethermalization in isolated quantum
  systems: a theoretical overview},
  \bibinfo{journal}{\href{http://dx.doi.org/10.1088/1361-6455/aabcdf}{J. Phys.
  B}} \href{http://dx.doi.org/10.1088/1361-6455/aabcdf}{{\bf
  \bibinfo{volume}{51}}, \bibinfo{pages}{112001}}
  (\href{http://dx.doi.org/10.1088/1361-6455/aabcdf}{\bibinfo{year}{2018}}).

\bibitem{deutsch_18}
\bibinfo{author}{J.~M. Deutsch}, \bibinfo{title}{Eigenstate thermalization
  hypothesis},
  \bibinfo{journal}{\href{http://dx.doi.org/10.1088/1361-6633/aac9f1}{Rep.
  Prog. Phys.}} \href{http://dx.doi.org/10.1088/1361-6633/aac9f1}{{\bf
  \bibinfo{volume}{81}}, \bibinfo{pages}{082001}}
  (\href{http://dx.doi.org/10.1088/1361-6633/aac9f1}{\bibinfo{year}{2018}}).

\bibitem{rigol_dunjko_08}
\bibinfo{author}{M.~Rigol}, \bibinfo{author}{V.~Dunjko}, and
  \bibinfo{author}{M.~Olshanii}, \bibinfo{title}{Thermalization and its
  mechanism for generic isolated quantum systems},
  \bibinfo{journal}{\href{http://dx.doi.org/10.1038/nature06838}{Nature
  (London)}} \href{http://dx.doi.org/10.1038/nature06838}{{\bf
  \bibinfo{volume}{452}}, \bibinfo{pages}{854}}
  (\href{http://dx.doi.org/10.1038/nature06838}{\bibinfo{year}{2008}}).

\bibitem{rigol_dunjko_07}
\bibinfo{author}{M.~Rigol}, \bibinfo{author}{V.~Dunjko},
  \bibinfo{author}{V.~Yurovsky}, and \bibinfo{author}{M.~Olshanii},
  \bibinfo{title}{Relaxation in a completely integrable many-body quantum
  system: An ab initio study of the dynamics of the highly excited states of
  {1D} lattice hard-core bosons},
  \bibinfo{journal}{\href{http://dx.doi.org/10.1103/PhysRevLett.98.050405}{Phys.
  Rev. Lett.}} \href{http://dx.doi.org/10.1103/PhysRevLett.98.050405}{{\bf
  \bibinfo{volume}{98}}, \bibinfo{pages}{050405}}
  (\href{http://dx.doi.org/10.1103/PhysRevLett.98.050405}{\bibinfo{year}{2007}}).

\bibitem{vidmar16}
\bibinfo{author}{L.~Vidmar} and \bibinfo{author}{M.~Rigol},
  \bibinfo{title}{{Generalized Gibbs ensemble in integrable lattice models}},
  \bibinfo{journal}{\href{http://dx.doi.org/10.1088/1742-5468/2016/06/064007}{J.
  Stat. Mech.}} \href{http://dx.doi.org/10.1088/1742-5468/2016/06/064007}{{\bf
  \bibinfo{volume}{{\rm (2016)}}}, \bibinfo{pages}{064007}}.

\bibitem{cazalilla_06}
\bibinfo{author}{M.~A. Cazalilla}, \bibinfo{title}{Effect of suddenly turning
  on interactions in the {Luttinger} model},
  \bibinfo{journal}{\href{http://dx.doi.org/10.1103/PhysRevLett.97.156403}{Phys.
  Rev. Lett.}} \href{http://dx.doi.org/10.1103/PhysRevLett.97.156403}{{\bf
  \bibinfo{volume}{97}}, \bibinfo{pages}{156403}}
  (\href{http://dx.doi.org/10.1103/PhysRevLett.97.156403}{\bibinfo{year}{2006}}).

\bibitem{rigol_muramatsu_06}
\bibinfo{author}{M.~Rigol}, \bibinfo{author}{A.~Muramatsu}, and
  \bibinfo{author}{M.~Olshanii}, \bibinfo{title}{Hard-core bosons on optical
  superlattices: {D}ynamics and relaxation in the superfluid and insulating
  regimes},
  \bibinfo{journal}{\href{http://dx.doi.org/10.1103/PhysRevA.74.053616}{Phys.
  Rev. A}} \href{http://dx.doi.org/10.1103/PhysRevA.74.053616}{{\bf
  \bibinfo{volume}{74}}, \bibinfo{pages}{053616}}
  (\href{http://dx.doi.org/10.1103/PhysRevA.74.053616}{\bibinfo{year}{2006}}).

\bibitem{iucci_cazalilla_09}
\bibinfo{author}{A.~Iucci} and \bibinfo{author}{M.~A. Cazalilla},
  \bibinfo{title}{Quantum quench dynamics of the {Luttinger} model},
  \bibinfo{journal}{\href{http://dx.doi.org/10.1103/PhysRevA.80.063619}{Phys.
  Rev. A}} \href{http://dx.doi.org/10.1103/PhysRevA.80.063619}{{\bf
  \bibinfo{volume}{80}}, \bibinfo{pages}{063619}}
  (\href{http://dx.doi.org/10.1103/PhysRevA.80.063619}{\bibinfo{year}{2009}}).

\bibitem{Cassidy_2011}
\bibinfo{author}{A.~C. Cassidy}, \bibinfo{author}{C.~W. Clark}, and
  \bibinfo{author}{M.~Rigol}, \bibinfo{title}{Generalized thermalization in an
  integrable lattice system},
  \bibinfo{journal}{\href{http://dx.doi.org/10.1103/physrevlett.106.140405}{Phys.
  Rev. Lett.}} \href{http://dx.doi.org/10.1103/physrevlett.106.140405}{{\bf
  \bibinfo{volume}{106}}, \bibinfo{pages}{140405}}
  (\href{http://dx.doi.org/10.1103/physrevlett.106.140405}{\bibinfo{year}{2011}}).

\bibitem{calabrese_essler_11}
\bibinfo{author}{P.~Calabrese}, \bibinfo{author}{F.~H.~L. Essler}, and
  \bibinfo{author}{M.~Fagotti}, \bibinfo{title}{Quantum quench in the
  transverse-field {I}sing chain},
  \bibinfo{journal}{\href{http://dx.doi.org/10.1103/PhysRevLett.106.227203}{Phys.
  Rev. Lett.}} \href{http://dx.doi.org/10.1103/PhysRevLett.106.227203}{{\bf
  \bibinfo{volume}{106}}, \bibinfo{pages}{227203}}
  (\href{http://dx.doi.org/10.1103/PhysRevLett.106.227203}{\bibinfo{year}{2011}}).

\bibitem{Gramsch_2012}
\bibinfo{author}{C.~Gramsch} and \bibinfo{author}{M.~Rigol},
  \bibinfo{title}{Quenches in a quasidisordered integrable lattice system:
  Dynamics and statistical description of observables after relaxation},
  \bibinfo{journal}{\href{http://dx.doi.org/10.1103/PhysRevA.86.053615}{Phys.
  Rev. A}} \href{http://dx.doi.org/10.1103/PhysRevA.86.053615}{{\bf
  \bibinfo{volume}{86}}, \bibinfo{pages}{053615}}
  (\href{http://dx.doi.org/10.1103/PhysRevA.86.053615}{\bibinfo{year}{2012}}).

\bibitem{calabrese_essler_12b}
\bibinfo{author}{P.~Calabrese}, \bibinfo{author}{F.~H.~L. Essler}, and
  \bibinfo{author}{M.~Fagotti}, \bibinfo{title}{Quantum quenches in the
  transverse field {I}sing chain: {II.} stationary state properties},
  \bibinfo{journal}{\href{http://dx.doi.org/10.1088/1742-5468/2012/07/P07022}{J.
  Stat. Mech.}} \href{http://dx.doi.org/10.1088/1742-5468/2012/07/P07022}{{\bf
  \bibinfo{volume}{2012}}, \bibinfo{pages}{P07022}}
  (\href{http://dx.doi.org/10.1088/1742-5468/2012/07/P07022}{\bibinfo{year}{2012}}).

\bibitem{essler_evangelisti_12}
\bibinfo{author}{F.~H.~L. Essler}, \bibinfo{author}{S.~Evangelisti}, and
  \bibinfo{author}{M.~Fagotti}, \bibinfo{title}{{Dynamical Correlations After a
  Quantum Quench}},
  \bibinfo{journal}{\href{http://dx.doi.org/10.1103/PhysRevLett.109.247206}{Phys.
  Rev. Lett.}} \href{http://dx.doi.org/10.1103/PhysRevLett.109.247206}{{\bf
  \bibinfo{volume}{109}}, \bibinfo{pages}{247206}}
  (\href{http://dx.doi.org/10.1103/PhysRevLett.109.247206}{\bibinfo{year}{2012}}).

\bibitem{caux_essler_13}
\bibinfo{author}{J.-S. Caux} and \bibinfo{author}{F.~H.~L. Essler},
  \bibinfo{title}{Time evolution of local observables after quenching to an
  integrable model},
  \bibinfo{journal}{\href{http://dx.doi.org/10.1103/PhysRevLett.110.257203}{Phys.
  Rev. Lett.}} \href{http://dx.doi.org/10.1103/PhysRevLett.110.257203}{{\bf
  \bibinfo{volume}{110}}, \bibinfo{pages}{257203}}
  (\href{http://dx.doi.org/10.1103/PhysRevLett.110.257203}{\bibinfo{year}{2013}}).

\bibitem{caux_konik_12}
\bibinfo{author}{J.-S. Caux} and \bibinfo{author}{R.~M. Konik},
  \bibinfo{title}{Constructing the generalized {G}ibbs ensemble after a quantum
  quench},
  \bibinfo{journal}{\href{http://dx.doi.org/10.1103/PhysRevLett.109.175301}{Phys.
  Rev. Lett.}} \href{http://dx.doi.org/10.1103/PhysRevLett.109.175301}{{\bf
  \bibinfo{volume}{109}}, \bibinfo{pages}{175301}}
  (\href{http://dx.doi.org/10.1103/PhysRevLett.109.175301}{\bibinfo{year}{2012}}).

\bibitem{kormos_shashi_13}
\bibinfo{author}{M.~Kormos}, \bibinfo{author}{A.~Shashi},
  \bibinfo{author}{Y.-Z. Chou}, \bibinfo{author}{J.-S. Caux}, and
  \bibinfo{author}{A.~Imambekov}, \bibinfo{title}{Interaction quenches in the
  one-dimensional {B}ose gas},
  \bibinfo{journal}{\href{http://dx.doi.org/10.1103/PhysRevB.88.205131}{Phys.
  Rev. B}} \href{http://dx.doi.org/10.1103/PhysRevB.88.205131}{{\bf
  \bibinfo{volume}{88}}, \bibinfo{pages}{205131}}
  (\href{http://dx.doi.org/10.1103/PhysRevB.88.205131}{\bibinfo{year}{2013}}).

\bibitem{pozsgay_13}
\bibinfo{author}{B.~Pozsgay}, \bibinfo{title}{The generalized {G}ibbs ensemble
  for {H}eisenberg spin chains},
  \bibinfo{journal}{\href{http://dx.doi.org/10.1088/1742-5468/2013/07/P07003}{J.
  Stat. Mech.}} \href{http://dx.doi.org/10.1088/1742-5468/2013/07/P07003}{{\bf
  \bibinfo{volume}{2013}}, \bibinfo{pages}{P07003}}
  (\href{http://dx.doi.org/10.1088/1742-5468/2013/07/P07003}{\bibinfo{year}{2013}}).

\bibitem{fagotti_collura_14}
\bibinfo{author}{M.~Fagotti}, \bibinfo{author}{M.~Collura},
  \bibinfo{author}{F.~H.~L. Essler}, and \bibinfo{author}{P.~Calabrese},
  \bibinfo{title}{Relaxation after quantum quenches in the spin-$\frac{1}{2}$
  {H}eisenberg {XXZ} chain},
  \bibinfo{journal}{\href{http://dx.doi.org/10.1103/PhysRevB.89.125101}{Phys.
  Rev. B}} \href{http://dx.doi.org/10.1103/PhysRevB.89.125101}{{\bf
  \bibinfo{volume}{89}}, \bibinfo{pages}{125101}}
  (\href{http://dx.doi.org/10.1103/PhysRevB.89.125101}{\bibinfo{year}{2014}}).

\bibitem{nardis_wouters_14}
\bibinfo{author}{J.~De~Nardis}, \bibinfo{author}{B.~Wouters},
  \bibinfo{author}{M.~Brockmann}, and \bibinfo{author}{J.-S. Caux},
  \bibinfo{title}{Solution for an interaction quench in the {Lieb-Liniger Bose}
  gas},
  \bibinfo{journal}{\href{http://dx.doi.org/10.1103/PhysRevA.89.033601}{Phys.
  Rev. A}} \href{http://dx.doi.org/10.1103/PhysRevA.89.033601}{{\bf
  \bibinfo{volume}{89}}, \bibinfo{pages}{033601}}
  (\href{http://dx.doi.org/10.1103/PhysRevA.89.033601}{\bibinfo{year}{2014}}).

\bibitem{wouters_denardis_14}
\bibinfo{author}{B.~Wouters}, \bibinfo{author}{J.~De~Nardis},
  \bibinfo{author}{M.~Brockmann}, \bibinfo{author}{D.~Fioretto},
  \bibinfo{author}{M.~Rigol}, and \bibinfo{author}{J.-S. Caux},
  \bibinfo{title}{Quenching the anisotropic {H}eisenberg chain: {E}xact
  solution and generalized {G}ibbs ensemble predictions},
  \bibinfo{journal}{\href{http://dx.doi.org/10.1103/PhysRevLett.113.117202}{Phys.
  Rev. Lett.}} \href{http://dx.doi.org/10.1103/PhysRevLett.113.117202}{{\bf
  \bibinfo{volume}{113}}, \bibinfo{pages}{117202}}
  (\href{http://dx.doi.org/10.1103/PhysRevLett.113.117202}{\bibinfo{year}{2014}}).

\bibitem{pozsgay_mestyan14}
\bibinfo{author}{B.~Pozsgay}, \bibinfo{author}{M.~Mesty\'an},
  \bibinfo{author}{M.~A. Werner}, \bibinfo{author}{M.~Kormos},
  \bibinfo{author}{G.~Zar\'and}, and \bibinfo{author}{G.~Tak\'acs},
  \bibinfo{title}{Correlations after quantum quenches in the {XXZ} spin chain:
  {F}ailure of the generalized {G}ibbs ensemble},
  \bibinfo{journal}{\href{http://dx.doi.org/10.1103/PhysRevLett.113.117203}{Phys.
  Rev. Lett.}} \href{http://dx.doi.org/10.1103/PhysRevLett.113.117203}{{\bf
  \bibinfo{volume}{113}}, \bibinfo{pages}{117203}}
  (\href{http://dx.doi.org/10.1103/PhysRevLett.113.117203}{\bibinfo{year}{2014}}).

\bibitem{Mierzejewski_2015}
\bibinfo{author}{M.~Mierzejewski}, \bibinfo{author}{P.~Prelov{\v{s}}ek}, and
  \bibinfo{author}{T.~Prosen}, \bibinfo{title}{Identifying local and quasilocal
  conserved quantities in integrable systems},
  \bibinfo{journal}{\href{http://dx.doi.org/10.1103/physrevlett.114.140601}{Phys.
  Rev. Lett.}} \href{http://dx.doi.org/10.1103/physrevlett.114.140601}{{\bf
  \bibinfo{volume}{114}}, \bibinfo{pages}{140601}}
  (\href{http://dx.doi.org/10.1103/physrevlett.114.140601}{\bibinfo{year}{2015}}).

\bibitem{ilievski_medenjak_15}
\bibinfo{author}{E.~Ilievski}, \bibinfo{author}{M.~Medenjak}, and
  \bibinfo{author}{T.~Prosen}, \bibinfo{title}{Quasilocal conserved operators
  in the isotropic {H}eisenberg spin-$1/2$ chain},
  \bibinfo{journal}{\href{http://dx.doi.org/10.1103/PhysRevLett.115.120601}{Phys.
  Rev. Lett.}} \href{http://dx.doi.org/10.1103/PhysRevLett.115.120601}{{\bf
  \bibinfo{volume}{115}}, \bibinfo{pages}{120601}}
  (\href{http://dx.doi.org/10.1103/PhysRevLett.115.120601}{\bibinfo{year}{2015}}).

\bibitem{ilievski_denardis_15}
\bibinfo{author}{E.~Ilievski}, \bibinfo{author}{J.~De~Nardis},
  \bibinfo{author}{B.~Wouters}, \bibinfo{author}{J.-S. Caux},
  \bibinfo{author}{F.~H.~L. Essler}, and \bibinfo{author}{T.~Prosen},
  \bibinfo{title}{Complete generalized {G}ibbs ensembles in an interacting
  theory},
  \bibinfo{journal}{\href{http://dx.doi.org/10.1103/PhysRevLett.115.157201}{Phys.
  Rev. Lett.}} \href{http://dx.doi.org/10.1103/PhysRevLett.115.157201}{{\bf
  \bibinfo{volume}{115}}, \bibinfo{pages}{157201}}
  (\href{http://dx.doi.org/10.1103/PhysRevLett.115.157201}{\bibinfo{year}{2015}}).

\bibitem{piroli_vernier_17}
\bibinfo{author}{L.~Piroli}, \bibinfo{author}{E.~Vernier},
  \bibinfo{author}{P.~Calabrese}, and \bibinfo{author}{M.~Rigol},
  \bibinfo{title}{Correlations and diagonal entropy after quantum quenches in
  {XXZ} chains},
  \bibinfo{journal}{\href{http://dx.doi.org/10.1103/PhysRevB.95.054308}{Phys.
  Rev. B}} \href{http://dx.doi.org/10.1103/PhysRevB.95.054308}{{\bf
  \bibinfo{volume}{95}}, \bibinfo{pages}{054308}}
  (\href{http://dx.doi.org/10.1103/PhysRevB.95.054308}{\bibinfo{year}{2017}}).

\bibitem{fagotti_maric_22}
\bibinfo{author}{M.~Fagotti}, \bibinfo{author}{V.~Mari\'{c}}, and
  \bibinfo{author}{L.~Zadnik}, \bibinfo{title}{{Nonequilibrium
  symmetry-protected topological order: emergence of semilocal Gibbs
  ensembles}},
  \href{https://arxiv.org/abs/2205.02221}{\bibinfo{howpublished}{arXiv:2205.02221}}.

\bibitem{calabrese_essler_review_16}
\bibinfo{author}{P.~Calabrese}, \bibinfo{author}{F.~H.~L. Essler}, and
  \bibinfo{author}{G.~Mussardo}, \bibinfo{title}{Introduction to `quantum
  integrability in out of equilibrium systems'},
  \bibinfo{journal}{\href{http://dx.doi.org/10.1088/1742-5468/2016/06/064001}{J.
  Stat. Mech.}} \href{http://dx.doi.org/10.1088/1742-5468/2016/06/064001}{{\bf
  \bibinfo{volume}{2016}}, \bibinfo{pages}{064001}}
  (\href{http://dx.doi.org/10.1088/1742-5468/2016/06/064001}{\bibinfo{year}{2016}}).

\bibitem{bertini2016transport}
\bibinfo{author}{B.~Bertini}, \bibinfo{author}{M.~Collura},
  \bibinfo{author}{J.~De~Nardis}, and \bibinfo{author}{M.~Fagotti},
  \bibinfo{title}{Transport in out-of-equilibrium {XXZ} chains: {E}xact
  profiles of charges and currents},
  \bibinfo{journal}{\href{http://dx.doi.org/10.1103/PhysRevLett.117.207201}{Phys.
  Rev. Lett.}} \href{http://dx.doi.org/10.1103/PhysRevLett.117.207201}{{\bf
  \bibinfo{volume}{117}}, \bibinfo{pages}{207201}}
  (\href{http://dx.doi.org/10.1103/PhysRevLett.117.207201}{\bibinfo{year}{2016}}).

\bibitem{castro2016emergent}
\bibinfo{author}{O.~A. Castro-Alvaredo}, \bibinfo{author}{B.~Doyon}, and
  \bibinfo{author}{T.~Yoshimura}, \bibinfo{title}{Emergent hydrodynamics in
  integrable quantum systems out of equilibrium},
  \bibinfo{journal}{\href{http://dx.doi.org/10.1103/PhysRevX.6.041065}{Phys.
  Rev. X}} \href{http://dx.doi.org/10.1103/PhysRevX.6.041065}{{\bf
  \bibinfo{volume}{6}}, \bibinfo{pages}{041065}}
  (\href{http://dx.doi.org/10.1103/PhysRevX.6.041065}{\bibinfo{year}{2016}}).

\bibitem{schemmer2019generalized}
\bibinfo{author}{M.~Schemmer}, \bibinfo{author}{I.~Bouchoule},
  \bibinfo{author}{B.~Doyon}, and \bibinfo{author}{J.~Dubail},
  \bibinfo{title}{Generalized hydrodynamics on an atom chip},
  \bibinfo{journal}{\href{http://dx.doi.org/10.1103/PhysRevLett.122.090601}{Phys.
  Rev. Lett.}} \href{http://dx.doi.org/10.1103/PhysRevLett.122.090601}{{\bf
  \bibinfo{volume}{122}}, \bibinfo{pages}{090601}}
  (\href{http://dx.doi.org/10.1103/PhysRevLett.122.090601}{\bibinfo{year}{2019}}).

\bibitem{malvania_zhang_21}
\bibinfo{author}{N.~Malvania}, \bibinfo{author}{Y.~Zhang},
  \bibinfo{author}{Y.~Le}, \bibinfo{author}{J.~Dubail},
  \bibinfo{author}{M.~Rigol}, and \bibinfo{author}{D.~S. Weiss},
  \bibinfo{title}{Generalized hydrodynamics in strongly interacting {1D B}ose
  gases},
  \bibinfo{journal}{\href{http://dx.doi.org/10.1126/science.abf0147}{Science}}
  \href{http://dx.doi.org/10.1126/science.abf0147}{{\bf \bibinfo{volume}{373}},
  \bibinfo{pages}{1129}}
  (\href{http://dx.doi.org/10.1126/science.abf0147}{\bibinfo{year}{2021}}).

\bibitem{Ziraldo_2012}
\bibinfo{author}{S.~Ziraldo}, \bibinfo{author}{A.~Silva}, and
  \bibinfo{author}{G.~E. Santoro}, \bibinfo{title}{Relaxation dynamics of
  disordered spin chains: Localization and the existence of a stationary
  state},
  \bibinfo{journal}{\href{http://dx.doi.org/10.1103/physrevlett.109.247205}{Phys.
  Rev. Lett.}} \href{http://dx.doi.org/10.1103/physrevlett.109.247205}{{\bf
  \bibinfo{volume}{109}}, \bibinfo{pages}{247205}}
  (\href{http://dx.doi.org/10.1103/physrevlett.109.247205}{\bibinfo{year}{2012}}).

\bibitem{Ziraldo_2013}
\bibinfo{author}{S.~Ziraldo} and \bibinfo{author}{G.~E. Santoro},
  \bibinfo{title}{Relaxation and thermalization after a quantum quench: Why
  localization is important},
  \bibinfo{journal}{\href{http://dx.doi.org/10.1103/physrevb.87.064201}{Phys.
  Rev. B}} \href{http://dx.doi.org/10.1103/physrevb.87.064201}{{\bf
  \bibinfo{volume}{87}}, \bibinfo{pages}{064201}}
  (\href{http://dx.doi.org/10.1103/physrevb.87.064201}{\bibinfo{year}{2013}}).

\bibitem{He_2013}
\bibinfo{author}{K.~He}, \bibinfo{author}{L.~F. Santos}, \bibinfo{author}{T.~M.
  Wright}, and \bibinfo{author}{M.~Rigol}, \bibinfo{title}{Single-particle and
  many-body analyses of a quasiperiodic integrable system after a quench},
  \bibinfo{journal}{\href{http://dx.doi.org/10.1103/physreva.87.063637}{Phys.
  Rev. A}} \href{http://dx.doi.org/10.1103/physreva.87.063637}{{\bf
  \bibinfo{volume}{87}}, \bibinfo{pages}{063637}}
  (\href{http://dx.doi.org/10.1103/physreva.87.063637}{\bibinfo{year}{2013}}).

\bibitem{wright_rigol_14}
\bibinfo{author}{T.~M. Wright}, \bibinfo{author}{M.~Rigol},
  \bibinfo{author}{M.~J. Davis}, and \bibinfo{author}{K.~V. Kheruntsyan},
  \bibinfo{title}{Nonequilibrium dynamics of one-dimensional hard-core anyons
  following a quench: Complete relaxation of one-body observables},
  \bibinfo{journal}{\href{http://dx.doi.org/10.1103/PhysRevLett.113.050601}{Phys.
  Rev. Lett.}} \href{http://dx.doi.org/10.1103/PhysRevLett.113.050601}{{\bf
  \bibinfo{volume}{113}}, \bibinfo{pages}{050601}}
  (\href{http://dx.doi.org/10.1103/PhysRevLett.113.050601}{\bibinfo{year}{2014}}).

\bibitem{Cramer_2008}
\bibinfo{author}{M.~Cramer}, \bibinfo{author}{C.~M. Dawson},
  \bibinfo{author}{J.~Eisert}, and \bibinfo{author}{T.~J. Osborne},
  \bibinfo{title}{Exact relaxation in a class of nonequilibrium quantum lattice
  systems},
  \bibinfo{journal}{\href{http://dx.doi.org/10.1103/physrevlett.100.030602}{Phys.
  Rev. Lett.}} \href{http://dx.doi.org/10.1103/physrevlett.100.030602}{{\bf
  \bibinfo{volume}{100}}}
  (\href{http://dx.doi.org/10.1103/physrevlett.100.030602}{\bibinfo{year}{2008}}).

\bibitem{Gluza_2016}
\bibinfo{author}{M.~Gluza}, \bibinfo{author}{C.~Krumnow},
  \bibinfo{author}{M.~Friesdorf}, \bibinfo{author}{C.~Gogolin}, and
  \bibinfo{author}{J.~Eisert}, \bibinfo{title}{Equilibration via
  {G}aussification in fermionic lattice systems},
  \bibinfo{journal}{\href{http://dx.doi.org/10.1103/physrevlett.117.190602}{Phys.
  Rev. Lett.}} \href{http://dx.doi.org/10.1103/physrevlett.117.190602}{{\bf
  \bibinfo{volume}{117}}}
  (\href{http://dx.doi.org/10.1103/physrevlett.117.190602}{\bibinfo{year}{2016}}).

\bibitem{murthy19}
\bibinfo{author}{C.~Murthy} and \bibinfo{author}{M.~Srednicki},
  \bibinfo{title}{Relaxation to gaussian and generalized gibbs states in
  systems of particles with quadratic hamiltonians},
  \bibinfo{journal}{\href{http://dx.doi.org/10.1103/PhysRevE.100.012146}{Phys.
  Rev. E}} \href{http://dx.doi.org/10.1103/PhysRevE.100.012146}{{\bf
  \bibinfo{volume}{100}}, \bibinfo{pages}{012146}}
  (\href{http://dx.doi.org/10.1103/PhysRevE.100.012146}{\bibinfo{year}{2019}}).

\bibitem{Gluza_2019}
\bibinfo{author}{M.~Gluza}, \bibinfo{author}{J.~Eisert}, and
  \bibinfo{author}{T.~Farrelly}, \bibinfo{title}{Equilibration towards
  generalized {G}ibbs ensembles in non-interacting theories},
  \bibinfo{journal}{\href{http://dx.doi.org/10.21468/scipostphys.7.3.038}{{SciPost}
  Physics}} \href{http://dx.doi.org/10.21468/scipostphys.7.3.038}{{\bf
  \bibinfo{volume}{7}}}
  (\href{http://dx.doi.org/10.21468/scipostphys.7.3.038}{\bibinfo{year}{2019}}).

\bibitem{lydzba_rigol_21}
\bibinfo{author}{P.~\L{}yd\ifmmode~\dot{z}\else \.{z}\fi{}ba},
  \bibinfo{author}{M.~Rigol}, and \bibinfo{author}{L.~Vidmar},
  \bibinfo{title}{{Entanglement in many-body eigenstates of quantum-chaotic
  quadratic Hamiltonians}},
  \bibinfo{journal}{\href{http://dx.doi.org/10.1103/PhysRevB.103.104206}{Phys.
  Rev. B}} \href{http://dx.doi.org/10.1103/PhysRevB.103.104206}{{\bf
  \bibinfo{volume}{103}}, \bibinfo{pages}{104206}}
  (\href{http://dx.doi.org/10.1103/PhysRevB.103.104206}{\bibinfo{year}{2021}}).

\bibitem{lydzba_zhang_21}
\bibinfo{author}{P.~\L{}yd\ifmmode~\dot{z}\else \.{z}\fi{}ba},
  \bibinfo{author}{Y.~Zhang}, \bibinfo{author}{M.~Rigol}, and
  \bibinfo{author}{L.~Vidmar}, \bibinfo{title}{{Single-particle eigenstate
  thermalization in quantum-chaotic quadratic Hamiltonians}},
  \bibinfo{journal}{\href{http://dx.doi.org/10.1103/PhysRevB.104.214203}{Phys.
  Rev. B}} \href{http://dx.doi.org/10.1103/PhysRevB.104.214203}{{\bf
  \bibinfo{volume}{104}}, \bibinfo{pages}{214203}}
  (\href{http://dx.doi.org/10.1103/PhysRevB.104.214203}{\bibinfo{year}{2021}}).

\bibitem{suntajs_prosen_21}
\bibinfo{author}{J.~\v{S}untajs}, \bibinfo{author}{T.~Prosen}, and
  \bibinfo{author}{L.~Vidmar}, \bibinfo{title}{{Spectral properties of
  three-dimensional Anderson model}},
  \bibinfo{journal}{\href{http://dx.doi.org/https://doi.org/10.1016/j.aop.2021.168469}{Ann.
  Phys.}}
  \href{http://dx.doi.org/https://doi.org/10.1016/j.aop.2021.168469}{{\bf
  \bibinfo{volume}{435}}, \bibinfo{pages}{168469}}
  (\href{http://dx.doi.org/https://doi.org/10.1016/j.aop.2021.168469}{\bibinfo{year}{2021}}).

\bibitem{ulcakar_vidmar_22}
\bibinfo{author}{I.~Ul\v{c}akar} and \bibinfo{author}{L.~Vidmar},
  \bibinfo{title}{Tight-binding billiards},
  \bibinfo{journal}{\href{http://dx.doi.org/10.1103/PhysRevE.106.034118}{Phys.
  Rev. E}} \href{http://dx.doi.org/10.1103/PhysRevE.106.034118}{{\bf
  \bibinfo{volume}{106}}, \bibinfo{pages}{034118}}
  (\href{http://dx.doi.org/10.1103/PhysRevE.106.034118}{\bibinfo{year}{2022}}).

\bibitem{lydzba_rigol_20}
\bibinfo{author}{P.~\L{}yd\ifmmode~\dot{z}\else \.{z}\fi{}ba},
  \bibinfo{author}{M.~Rigol}, and \bibinfo{author}{L.~Vidmar},
  \bibinfo{title}{Eigenstate entanglement entropy in random quadratic
  {Hamiltonians}},
  \bibinfo{journal}{\href{http://dx.doi.org/10.1103/PhysRevLett.125.180604}{Phys.
  Rev. Lett.}} \href{http://dx.doi.org/10.1103/PhysRevLett.125.180604}{{\bf
  \bibinfo{volume}{125}}, \bibinfo{pages}{180604}}
  (\href{http://dx.doi.org/10.1103/PhysRevLett.125.180604}{\bibinfo{year}{2020}}).

\bibitem{liu_chen_18}
\bibinfo{author}{C.~Liu}, \bibinfo{author}{X.~Chen}, and
  \bibinfo{author}{L.~Balents}, \bibinfo{title}{Quantum entanglement of the
  {Sachdev-Ye-Kitaev} models},
  \bibinfo{journal}{\href{http://dx.doi.org/https://doi.org/10.1103/PhysRevB.97.245126}{Phys.
  Rev. B}}
  \href{http://dx.doi.org/https://doi.org/10.1103/PhysRevB.97.245126}{{\bf
  \bibinfo{volume}{97}}, \bibinfo{pages}{245126}}
  (\href{http://dx.doi.org/https://doi.org/10.1103/PhysRevB.97.245126}{\bibinfo{year}{2018}}).

\bibitem{altshuler_shklovskii_86}
\bibinfo{author}{B.~Al'tshuler} and \bibinfo{author}{B.~Shklovskii},
  \bibinfo{title}{Repulsion of energy levels and conductivity of small metal
  sample}, \bibinfo{journal}{Zh. Eksp. Teor. Fiz.} {\bf \bibinfo{volume}{91}},
  \bibinfo{pages}{220}  (\bibinfo{year}{1986}).

\bibitem{altshuler_zharekeshev_88}
\bibinfo{author}{B.~Al'tshuler}, \bibinfo{author}{I.~Zharekeshev},
  \bibinfo{author}{S.~Kotochigova}, and \bibinfo{author}{B.~Shklovskii},
  \bibinfo{title}{Repulsion between energy levels and the metal-insulator
  transition}, \bibinfo{journal}{Zh. Eksp. Teor. Fiz.} {\bf
  \bibinfo{volume}{94}}, \bibinfo{pages}{343}  (\bibinfo{year}{1988}).

\bibitem{Note1}
\bibinfo{note}{Namely, for lattice systems such as the ones of interest here,
  satisfying $\protect \frac {1}{V}{\protect \rm Tr}\{ \protect \hat {\protect
  \cal O}^2 \}=1$, where $V$ is the number of lattice sites.}

\bibitem{srednicki_99}
\bibinfo{author}{M.~Srednicki}, \bibinfo{title}{The approach to thermal
  equilibrium in quantized chaotic systems},
  \bibinfo{journal}{\href{http://dx.doi.org/10.1088/0305-4470/32/7/007}{J.
  Phys. A.}} \href{http://dx.doi.org/10.1088/0305-4470/32/7/007}{{\bf
  \bibinfo{volume}{32}}, \bibinfo{pages}{1163}}
  (\href{http://dx.doi.org/10.1088/0305-4470/32/7/007}{\bibinfo{year}{1999}}).

\bibitem{bianchi_hackal_21}
\bibinfo{author}{E.~Bianchi}, \bibinfo{author}{L.~Hackl}, and
  \bibinfo{author}{M.~Kieburg}, \bibinfo{title}{Page curve for fermionic
  {G}aussian states},
  \bibinfo{journal}{\href{http://dx.doi.org/10.1103/PhysRevB.103.L241118}{Phys.
  Rev. B}} \href{http://dx.doi.org/10.1103/PhysRevB.103.L241118}{{\bf
  \bibinfo{volume}{103}}, \bibinfo{pages}{L241118}}
  (\href{http://dx.doi.org/10.1103/PhysRevB.103.L241118}{\bibinfo{year}{2021}}).

\bibitem{bianchi_hackl_22}
\bibinfo{author}{E.~Bianchi}, \bibinfo{author}{L.~Hackl},
  \bibinfo{author}{M.~Kieburg}, \bibinfo{author}{M.~Rigol}, and
  \bibinfo{author}{L.~Vidmar}, \bibinfo{title}{{Volume-Law Entanglement Entropy
  of Typical Pure Quantum States}},
  \bibinfo{journal}{\href{http://dx.doi.org/10.1103/PRXQuantum.3.030201}{PRX
  Quantum}} \href{http://dx.doi.org/10.1103/PRXQuantum.3.030201}{{\bf
  \bibinfo{volume}{3}}, \bibinfo{pages}{030201}}
  (\href{http://dx.doi.org/10.1103/PRXQuantum.3.030201}{\bibinfo{year}{2022}}).

\bibitem{Projected_1}
\bibinfo{author}{M.~Lucas}, \bibinfo{author}{L.~Piroli},
  \bibinfo{author}{J.~De~Nardis}, and \bibinfo{author}{A.~De~Luca},
  \bibinfo{title}{{Generalized deep thermalization for free fermions}},
  \href{https://arxiv.org/abs/2207.13628}{\bibinfo{howpublished}{arXiv:2207.13628}}.

\bibitem{Note2}
\bibinfo{note}{Namely, observables that have an ${\protect \bf O}(1)$ number of
  nondegenerate eigenvalues in the single-particle spectrum.}

\bibitem{Venuti_13}
\bibinfo{author}{L.~C. Venuti} and \bibinfo{author}{P.~Zanardi},
  \bibinfo{title}{Gaussian equilibration},
  \bibinfo{journal}{\href{http://dx.doi.org/10.1103/PhysRevE.87.012106}{Phys.
  Rev. E}} \href{http://dx.doi.org/10.1103/PhysRevE.87.012106}{{\bf
  \bibinfo{volume}{87}}, \bibinfo{pages}{012106}}
  (\href{http://dx.doi.org/10.1103/PhysRevE.87.012106}{\bibinfo{year}{2013}}).

\bibitem{Venuti_2015}
\bibinfo{author}{L.~C. Venuti}, \bibinfo{title}{Theory of temporal fluctuations
  in isolated quantum systems}, {\em \bibinfo{booktitle}{Quantum Criticality in
  Condensed Matter}\/} (\bibinfo{publisher}{{WORLD} {SCIENTIFIC}},
  \bibinfo{year}{2015}).

\bibitem{suppmat}
\bibinfo{note}{See {Supplemental Material} for the derivation of
  Eq.~(\ref{eq_bound1}), equilibration of $q$-body observables, absence of ETH
  in many-body energy eigenstates, details on the GGE in the SYK2 model, the
  difference between the GGE and GE in the 3D Anderson model, and the
  smoothness of Lagrange multipliers.}

\bibitem{Anderson_58}
\bibinfo{author}{P.~W. Anderson}, \bibinfo{title}{Absence of diffusion in
  certain random lattices},
  \bibinfo{journal}{\href{http://dx.doi.org/10.1103/PhysRev.109.1492}{Phys.
  Rev.}} \href{http://dx.doi.org/10.1103/PhysRev.109.1492}{{\bf
  \bibinfo{volume}{109}}, \bibinfo{pages}{1492}}
  (\href{http://dx.doi.org/10.1103/PhysRev.109.1492}{\bibinfo{year}{1958}}).

\bibitem{Slevin_2014}
\bibinfo{author}{K.~Slevin} and \bibinfo{author}{T.~Ohtsuki},
  \bibinfo{title}{Critical exponent for the anderson transition in the
  three-dimensional orthogonal universality class},
  \bibinfo{journal}{\href{http://dx.doi.org/10.1088/1367-2630/16/1/015012}{New
  Journal of Physics}}
  \href{http://dx.doi.org/10.1088/1367-2630/16/1/015012}{{\bf
  \bibinfo{volume}{16}}, \bibinfo{pages}{015012}}
  (\href{http://dx.doi.org/10.1088/1367-2630/16/1/015012}{\bibinfo{year}{2014}}).

\bibitem{slevin_ohtsuki_18}
\bibinfo{author}{K.~Slevin} and \bibinfo{author}{T.~Ohtsuki},
  \bibinfo{title}{Critical exponent of the {Anderson} transition using
  massively parallel supercomputing},
  \bibinfo{journal}{\href{http://dx.doi.org/10.7566/JPSJ.87.094703}{J. Phys.
  Soc. Jpn.}} \href{http://dx.doi.org/10.7566/JPSJ.87.094703}{{\bf
  \bibinfo{volume}{87}}, \bibinfo{pages}{094703}}
  (\href{http://dx.doi.org/10.7566/JPSJ.87.094703}{\bibinfo{year}{2018}}).

\bibitem{Kinoshita2006}
\bibinfo{author}{T.~Kinoshita}, \bibinfo{author}{T.~Wenger}, and
  \bibinfo{author}{S.~D.~Weiss}, \bibinfo{title}{{A quantum Newton's cradle}},
  \bibinfo{journal}{\href{http://dx.doi.org/10.1038/nature04693}{Nature
  (London)}} \href{http://dx.doi.org/10.1038/nature04693}{{\bf
  \bibinfo{volume}{440}}, \bibinfo{pages}{900}}
  (\href{http://dx.doi.org/10.1038/nature04693}{\bibinfo{year}{2006}}).

\bibitem{Sachdev_93}
\bibinfo{author}{S.~Sachdev} and \bibinfo{author}{J.~Ye},
  \bibinfo{title}{Gapless spin-fluid ground state in a random quantum
  heisenberg magnet},
  \bibinfo{journal}{\href{http://dx.doi.org/10.1103/PhysRevLett.70.3339}{Phys.
  Rev. Lett.}} \href{http://dx.doi.org/10.1103/PhysRevLett.70.3339}{{\bf
  \bibinfo{volume}{70}}, \bibinfo{pages}{3339}}
  (\href{http://dx.doi.org/10.1103/PhysRevLett.70.3339}{\bibinfo{year}{1993}}).

\bibitem{Cazalilla_2012}
\bibinfo{author}{M.~A. Cazalilla}, \bibinfo{author}{A.~Iucci}, and
  \bibinfo{author}{M.-C. Chung}, \bibinfo{title}{Thermalization and quantum
  correlations in exactly solvable models},
  \bibinfo{journal}{\href{http://dx.doi.org/10.1103/PhysRevE.85.011133}{Phys.
  Rev. E}} \href{http://dx.doi.org/10.1103/PhysRevE.85.011133}{{\bf
  \bibinfo{volume}{85}}, \bibinfo{pages}{011133}}
  (\href{http://dx.doi.org/10.1103/PhysRevE.85.011133}{\bibinfo{year}{2012}}).

\bibitem{Kita_2015}
\bibinfo{author}{T.~Kita}, {\em \bibinfo{title}{Density Matrices and
  Two-Particle Correlations}\/}, {\em \bibinfo{booktitle}{Statistical Mechanics
  of Superconductivity}\/}, \bibinfo{pages}{61--71}
  (\bibinfo{publisher}{Springer Japan}, \bibinfo{address}{Tokyo},
  \bibinfo{year}{2015}).

\bibitem{Yang_1962}
\bibinfo{author}{C.~N. Yang}, \bibinfo{title}{Concept of off-diagonal
  long-range order and the quantum phases of liquid he and of superconductors},
  \bibinfo{journal}{\href{http://dx.doi.org/10.1103/RevModPhys.34.694}{Rev.
  Mod. Phys.}} \href{http://dx.doi.org/10.1103/RevModPhys.34.694}{{\bf
  \bibinfo{volume}{34}}, \bibinfo{pages}{694}}
  (\href{http://dx.doi.org/10.1103/RevModPhys.34.694}{\bibinfo{year}{1962}}).

\bibitem{haque_mcclarty_19}
\bibinfo{author}{M.~Haque} and \bibinfo{author}{P.~A. McClarty},
  \bibinfo{title}{{Eigenstate thermalization scaling in Majorana clusters: From
  chaotic to integrable Sachdev-Ye-Kitaev models}},
  \bibinfo{journal}{\href{http://dx.doi.org/10.1103/PhysRevB.100.115122}{Phys.
  Rev. B}} \href{http://dx.doi.org/10.1103/PhysRevB.100.115122}{{\bf
  \bibinfo{volume}{100}}, \bibinfo{pages}{115122}}
  (\href{http://dx.doi.org/10.1103/PhysRevB.100.115122}{\bibinfo{year}{2019}}).

\bibitem{Fitz_2011}
\bibinfo{author}{M.~Rigol} and \bibinfo{author}{M.~Fitzpatrick},
  \bibinfo{title}{Initial-state dependence of the quench dynamics in integrable
  quantum systems},
  \bibinfo{journal}{\href{http://dx.doi.org/10.1103/PhysRevA.84.033640}{Phys.
  Rev. A}} \href{http://dx.doi.org/10.1103/PhysRevA.84.033640}{{\bf
  \bibinfo{volume}{84}}, \bibinfo{pages}{033640}}
  (\href{http://dx.doi.org/10.1103/PhysRevA.84.033640}{\bibinfo{year}{2011}}).

\bibitem{He_2012}
\bibinfo{author}{K.~He} and \bibinfo{author}{M.~Rigol},
  \bibinfo{title}{Initial-state dependence of the quench dynamics in integrable
  quantum systems. ii. thermal states},
  \bibinfo{journal}{\href{http://dx.doi.org/10.1103/PhysRevA.85.063609}{Phys.
  Rev. A}} \href{http://dx.doi.org/10.1103/PhysRevA.85.063609}{{\bf
  \bibinfo{volume}{85}}, \bibinfo{pages}{063609}}
  (\href{http://dx.doi.org/10.1103/PhysRevA.85.063609}{\bibinfo{year}{2012}}).

\bibitem{He_2013b}
\bibinfo{author}{K.~He} and \bibinfo{author}{M.~Rigol},
  \bibinfo{title}{Initial-state dependence of the quench dynamics in integrable
  quantum systems. iii. chaotic states},
  \bibinfo{journal}{\href{http://dx.doi.org/10.1103/PhysRevA.87.043615}{Phys.
  Rev. A}} \href{http://dx.doi.org/10.1103/PhysRevA.87.043615}{{\bf
  \bibinfo{volume}{87}}, \bibinfo{pages}{043615}}
  (\href{http://dx.doi.org/10.1103/PhysRevA.87.043615}{\bibinfo{year}{2013}}).

\end{thebibliography}


\newpage
\phantom{a}
\newpage
\setcounter{figure}{0}
\setcounter{equation}{0}
\setcounter{table}{0}

\renewcommand{\thetable}{S\arabic{table}}
\renewcommand{\thefigure}{S\arabic{figure}}
\renewcommand{\theequation}{S\arabic{equation}}
\renewcommand{\thepage}{S\arabic{page}}

\renewcommand{\thesection}{S\arabic{section}}

\onecolumngrid

\begin{center}

{\large \bf Supplemental Material:\\
Generalized Thermalization in Quantum-Chaotic Quadratic Hamiltonians}\\

\vspace{0.3cm}

\setcounter{page}{1}

Patrycja  \L yd\.{z}ba$^{1}$,  Marcin Mierzejewski$^{1}$, Marcos Rigol$^{2}$ and Lev Vidmar$^{3,4}$\\
$^1${\it Department of Theoretical Physics, Wroclaw University of Science and Technology, 50-370 Wroc{\l}aw, Poland}
$^2${\it Department of Physics, The Pennsylvania State University, University Park, Pennsylvania 16802, USA}
$^3${\it Department of Theoretical Physics, J. Stefan Institute, SI-1000 Ljubljana, Slovenia} \\
$^4${\it Department of Physics, Faculty of Mathematics and Physics, University of Ljubljana, SI-1000 Ljubljana, Slovenia}

\end{center}

\vspace{0.6cm}

\twocolumngrid

\label{pagesupp}

\section{Derivation of Eq.~(\ref{eq_bound1})}

The expectation value of a one-body observable $\hat{O}=\sum_{\alpha\beta} O_{\alpha\beta} \hat{f}_\alpha^\dagger \hat{f}_\beta$ in a quadratic model after a quantum quench can be written as
\begin{equation}
     \langle\hat{O}(t)\rangle = \sum_{\alpha,\beta=1}^V R_{\alpha\beta} O_{\beta\alpha} e^{i(\epsilon_\beta - \epsilon_\alpha)t} \;,
\end{equation}
where $|\alpha\rangle$ and $|\beta\rangle$ are single-particle eigenstates of the final Hamiltonian, with single-particle eigenenergies $\epsilon_\alpha$ and $\epsilon_\beta$, respectively. $R_{\alpha\beta} = \langle\Psi_0|\hat{f}_{\beta}^\dagger \hat{f}^{}_{\alpha}|\Psi_0\rangle$ is the one-body correlation matrix of the initial state $|\Psi_0\rangle$, and $O_{\alpha\beta} = \langle \alpha| \hat{O} |\beta\rangle$ are the matrix elements of the one-body observable in the single-particle Hilbert space. 

The variance of the temporal fluctuations of $\langle\hat{O}(t)\rangle$ is
\begin{equation}\label{eqtemp}
\sigma_t^2 = \overline{\langle\hat{O}(t)\rangle^2} - \overline{\langle\hat{O}(t)\rangle}^2 \;,
\end{equation}
where $\overline{\langle\hat{O}(t)\rangle} = \lim_{\tau\to\infty} \frac{1}{\tau} \int_{0}^{\tau} \langle\hat{O}(t)\rangle dt$. 

The infinite-time average in the first term of Eq.~(\ref{eqtemp}) can be written as
\begin{equation}
\begin{split}
    \overline{\langle\hat{O}(t)\rangle^2} = 
    & \sum_{\alpha,\beta,\omega,\rho=1}^{V} O_{\beta\alpha}O_{\rho\omega}
    R_{\alpha\beta} R_{\omega\rho}\times\\
    &\lim_{\tau\to\infty} \frac{1}{\tau}\int_{0}^{\tau} e^{i\left(\epsilon_\beta - \epsilon_\alpha + \epsilon_\rho -\epsilon_\omega\right)t}\;dt\;.
\end{split}
\end{equation}
Assuming that energy gaps are not degenerated in the single-particle spectrum, the integrand is equal to one for $\alpha=\beta,\, \omega=\rho$ and $\alpha=\rho \neq \beta=\omega$, so we can write
\begin{equation}\label{eq:squareo}
    \overline{\langle\hat{O}(t)\rangle^2}=\sum_{\alpha,\beta=1}^{V} O_{\alpha\alpha} O_{\beta\beta} R_{\alpha\alpha} R_{\beta\beta} + \sum_{\alpha\neq\beta=1}^{V} |O_{\alpha\beta}|^2 |R_{\alpha\beta}|^2\;.
\end{equation}

The infinite-time average in the second term of Eq.~(\ref{eqtemp}) can be written as
\begin{equation}
\overline{\langle\hat{O}(t)\rangle} =
\sum_{\alpha,\beta=1}^{V} O_{\beta\alpha} R_{\alpha\beta} \lim_{\tau\to\infty} \frac{1}{\tau} \int_{0}^{\tau} e^{i\left(\epsilon_\beta-\epsilon_\alpha\right)t}\;dt\;.
\end{equation}
Assuming that there are no degeneracies in the single-particle spectrum, one can write
\begin{equation}
    \overline{\langle\hat{O}(t)\rangle}=\sum_{\alpha=1}^{V} O_{\alpha\alpha}R_{\alpha\alpha}
\end{equation}
so that
\begin{equation}
\label{eq:osquare}
\overline{\langle\hat{O}(t)\rangle}^2=\sum_{\alpha,\beta=1}^{V} O_{\alpha\alpha}O_{\beta\beta}R_{\alpha\alpha}R_{\beta\beta}\;.
\end{equation}
Substituting Eqs.~\eqref{eq:squareo} and~\eqref{eq:osquare} in Eq.~\eqref{eqtemp}, one obtains the variance of temporal fluctuations from Eq.~(\ref{eq_bound1})
\begin{equation}
    \sigma_t^2 =\sum_{\alpha\ne\beta=1}^V |O_{\alpha\beta}|^2 |R_{\alpha\beta}|^2\,.
\end{equation}

\section{Equilibration of $q$-body observables}

In this section, we generalize the proof of equilibration of one-body observables to $q$-body observables with an arbitrary finite $q$. 
Using Wick's theorem~\cite{Cazalilla_2012, Kita_2015}, expectation values of $q$-body ($q=2,\,3,\,\ldots$) operators can be expressed as finite sums of products of expectation values of one-body operators.
If the temporal fluctuations of all one-body operators involved vanish in the thermodynamic limit, then the $q$-body operator equilibrates to the GGE prediction~\cite{Ziraldo_2012, Ziraldo_2013}, i.e., it exhibits generalized thermalization.

The purpose of this section is to generalize our proof of equilibration of one-body observables to $q$-body observables for arbitrary initial states, i.e., without relying on Wick's theorem.
The generalization is valid for $q$-body observables $\hat{O}(q)$ that can be written as products of $q$ one-body observables $\hat{O}(1)^{(i)}$, $i\in\{1,...,q\}$. Namely, for observables that can be written as
    \begin{equation}
    \hat{O}(q)=\sum\limits_{\substack{\alpha_1,...,\alpha_q \\ \beta_1, ..., \beta_q}}O(1)^{(1)}_{\alpha_1\beta_1} ...\;O(1)^{(q)}_{\alpha_q\beta_q} \hat{f}_{\alpha_1}^\dagger \hat{f}_{\beta_1} ...\;\hat{f}_{\alpha_q}^\dagger \hat{f}_{\beta_q},
    \end{equation}
    where $\alpha_{i},\beta_{i}\in\left\{1,...,V\right\}$.
Our derivations are based on the following assumptions and properties:

\begin{enumerate}[wide, labelwidth=!, labelindent=0pt]
    
    \item The one-body observables $\hat{O}(1)^{(i)}$ exhibit single-particle eigenstate thermalization.
    For simplicity, we assume the matrix elements to be structureless, so that $\text{max}_{\alpha,\beta}\left\{|O(1)^{(i)}_{\alpha\beta} |\right\} = \textbf{O}(1/V)$.

    \item The $q$-body density matrix is defined as
    \begin{equation}
    \hat{R}(q)=\sum\limits_{\substack{\alpha_1,...,\alpha_q \\ \beta_1, ..., \beta_q}} R(q)_{\alpha_1...\alpha_q}^{\beta_1...\beta_q} \hat{f}_{\alpha_1}^\dagger ... \hat{f}_{\alpha_q}^\dagger \hat{f}_{\beta_q} ... \hat{f}_{\beta_1}
    \end{equation}
    with 
    \begin{equation}
    R(q)_{\alpha_1...\alpha_q}^{\beta_1...\beta_q}=\langle\Psi_0| \hat{f}_{\beta_1}^\dagger ... \hat{f}_{\beta_q}^\dagger \hat{f}_{\alpha_q} ... \hat{f}_{\alpha_1} |\Psi_0\rangle.
    \end{equation}
    Permutations of $\alpha_{i}$ (or $\beta_{i}$) can only change the sign of $R(q)_{\alpha_1...\alpha_q}^{\beta_1...\beta_q}$.
    The partial trace of ${R}(q)$ simplifies to
    \begin{equation} \label{eq:R_partial_trace}
    \sum_{\alpha}R(q)_{\alpha\alpha_2...\alpha_q}^{\alpha\beta_2...\beta_q}=\mathbf{O}(V) R(q-1)_{\alpha_2...\alpha_q}^{\beta_2...\beta_q}.
    \end{equation}
    While in Eq.~(\ref{eq:R_partial_trace}) the sum is done over the first indices, it can equally be done over an arbitrary pair of indices $\alpha=\alpha_{i}=\beta_{j}$. The constant $\mathbf{O}(V)$ equals $\pm(N-q+1)$, where $N$ is the particle number.
    Note that the one-body density matrix $\hat{R}$ with elements $R_{\alpha\beta}$, introduced in the main text, corresponds to $\hat{R}(1)$ with elements $R(1)_{\alpha}^{\beta}$.
    
    \item The trace of ${R}(q)$ simplifies to 
    \begin{equation}\label{eq:traceq}
    {\rm Tr} \left[R(q)\right]=\sum\limits_{\alpha_1,...,\alpha_q} R(q)_{\alpha_1...\alpha_q}^{\alpha_1...\alpha_q}=\mathbf{O}(V^q),
    \end{equation}
    for an arbitrary permutation of upper and lower indices. The constant $\mathbf{O}(V^q)$ equals $\pm N!/(N-q)!$. Simultaneously, the maximal eigenvalue of ${R}(q)$ increases with the system size as~\cite{Yang_1962}
    \begin{equation}\label{eq:maxeigq}
        \lambda_\text{max}(q)\le
        \left\{
        \begin{array}{ll}
        \textbf{O}(V^{q/2}) & \text{ for }q = \text{even},\\ 
        \textbf{O}(V^{(q-1)/2}) & \text{ for }q = \text{odd}.
        \end{array}
        \right.
    \end{equation}
    One can show that Eqs.~\eqref{eq:traceq} and~\eqref{eq:maxeigq} lead to 
    \begin{equation}
        {\rm Tr}\left[R(q)^2\right]\le
        \left\{
        \begin{array}{ll}
        \textbf{O}(V^{3q/2}) & \text{ for }q = \text{even},\\ 
        \textbf{O}(V^{(3q-1)/2}) & \text{ for }q = \text{odd}.
        \end{array}
        \right.
    \end{equation}
    Our derivations are based primarily on this result.
    
    \item The standard deviation satisfies the inequality 
    \begin{equation}
    \sigma_t(\hat{A} + \hat{B}) \le \sigma_t(\hat{A}) + \sigma_t(\hat{B}),
    \end{equation}
    where $\sigma_t^2(\hat{A})=\overline{\langle\hat{A}(t)\rangle^2}-\overline{\langle\hat{A}(t)\rangle}^2$. 
    
\end{enumerate}

In Sec.~\ref{sec:two-body}, we show how the assumptions and properties above ensure the equilibration of two-body observables. In Sec.~\ref{sec:q-body}, we discuss the general case of $q$-body observables.

\subsection{Two-body observables} \label{sec:two-body}

Let us consider an observable
\begin{equation}
    \hat{O}(2) = \sum\limits_{\substack{\alpha_1,\alpha_2 \\ \beta_1,\beta_2}} O(1)^{(1)}_{\alpha_1\beta_1}O(1)^{(2)}_{\alpha_2\beta_2} \hat{f}_{\alpha_1}^\dagger \hat{f}_{\beta_1}\hat{f}_{\alpha_2}^\dagger \hat{f}_{\beta_2}.
\end{equation}
If we move all creation operators to the left, we arrive at 
\begin{equation}
\begin{split}
    \hat{O}(2) = & \sum_{\alpha_1,\beta_2}\left(\sum_{\beta_1} O(1)^{(1)}_{\alpha_1\beta_1}O(1)^{(2)}_{\beta_1\beta_2} \right)\hat{f}_{\alpha_1}^\dagger\hat{f}_{\beta_2}\\
    & + \sum\limits_{\substack{\alpha_1,\alpha_2 \\ \beta_1,\beta_2}} O(1)^{(1)}_{\alpha_1\beta_1}O(1)^{(2)}_{\alpha_2\beta_2} \hat{f}_{\alpha_1}^\dagger \hat{f}_{\alpha_2}^\dagger \hat{f}_{\beta_2} \hat{f}_{\beta_1}.
\end{split}
\end{equation}
If matrices ${O}(1)^{(1)}_{\alpha_1,\beta_1}$ and ${O}(1)^{(2)}_{\beta_1,\beta_2}$ satisfy the assumption 1, their product also satisfies this assumption, since 
\begin{equation}
\begin{split}
    & \text{max}_{\alpha_1,\beta_2}\left\{\left|\left[{O}(1)^{(1)}{O}(1)^{(2)}\right]_{\alpha_1\beta_2}\right|\right\} = \\
    & \text{max}_{\alpha_1,\beta_2}\left\{\left|\sum_{\beta_1} O(1)^{(1)}_{\alpha_1\beta_1}O(1)^{(2)}_{\beta_1\beta_2}\right|\right\} \le\\
    & V \; \left(\text{max}_{\alpha,\beta,i}\left\{\left|O(1)^{(i)}_{\alpha\beta}\right|\right\} \right)^2 = \textbf{O}(1/V).
\end{split}
\end{equation}
Thus, we can write $\hat{O}(2)$ as a sum of $k$-body observables $\hat{\mathcal{O}}(k)^{(k)}$ with $k\in\left\{1,2\right\}$
\begin{equation}
    \hat{O}(2)=\sum_{k=1}^{2} \hat{\mathcal{O}}\left(k\right)^{(k)},
\end{equation}
where
\begin{equation}
    \hat{\mathcal{O}}(1)^{(1)}=\sum\limits_{\substack{\alpha_1 \beta_1}} \tilde{\mathcal{O}}(1)^{(1,1)}_{\alpha_1\beta_1}\hat{f}_{\alpha_1}^\dagger\hat{f}_{\beta_1},
\end{equation}
and
\begin{equation}
    \hat{\mathcal{O}}(2)^{(2)}=\sum\limits_{\substack{\alpha_1 \alpha_2 \\ \beta_1 \beta_2}} \tilde{\mathcal{O}}(1)^{(2,1)}_{\alpha_1\beta_1} \tilde{\mathcal{O}}(1)^{(2,2)}_{\alpha_2\beta_2}\hat{f}_{\alpha_1}^\dagger\hat{f}_{\alpha_2}^\dagger\hat{f}_{\beta_2}\hat{f}_{\beta_1}.
\end{equation}
The matrices $\tilde{\mathcal{O}}(1)^{(k,i)}_{\alpha_1\beta_1}$ satisfy the assumption 1.

Thanks to property 4, we can consider the standard deviations of operators $\hat{\mathcal{O}}(1)^{(1)}$ and $\hat{\mathcal{O}}(2)^{(2)}$ separately. In the main text, we computed the standard deviation of the former, $\sigma_t(\hat{\mathcal{O}}(1)^{(1)})\le\textbf{O}(V^{-1/2})$, so next we focus on the standard deviation of the latter,
\begin{equation}
    \sigma_t(\hat{\mathcal{O}}(2))=\sqrt{\overline{\langle\hat{\mathcal{O}}(2)\rangle_t^2}-\overline{\langle\hat{\mathcal{O}}(2)\rangle}_t^2}.
\end{equation}
Note that we have simplified the notation $\langle\hat{\mathcal{O}}(2)^{(2)}(t)\rangle\rightarrow\langle\hat{\mathcal{O}}(2)\rangle_t$. In what follows, we drop the upper index $k=2$ in the observable $\hat{\mathcal{O}}(2)^{(2)}\rightarrow \hat{\mathcal{O}}(2)$ and the matrices $\tilde{\mathcal{O}}_{\alpha_1,\beta_1}^{(2,i)}\rightarrow \tilde{\mathcal{O}}_{\alpha_1,\beta_1}^{(i)}$.

The expectation value of $\hat{\mathcal{O}}(2)$ at time $t$ is given by
\begin{equation}
     \langle\hat{\mathcal{O}}(2)\rangle_t = \sum_{\Omega}\langle\Omega|e^{-i\hat{H}t}\hat{\rho}_0e^{i\hat{H}t}\hat{\mathcal{O}}(2)|\Omega\rangle,
\end{equation}
where $|\Omega\rangle$ are many-body energy eigenstates and $\hat{\rho}_0=|\Psi_0\rangle\langle\Psi_0|$. Therefore
\begin{equation}
    \langle\hat{\mathcal{O}}(2)\rangle_t\!=\hspace{-0.2cm}\sum\limits_{\substack{\alpha_1,\alpha_2 \\ \beta_1,\beta_2}}\! \tilde{\mathcal{O}}(1)^{(1)}_{\alpha_1\beta_1}\tilde{\mathcal{O}}(1)^{(2)}_{\alpha_2\beta_2}R(2)_{\beta_1\beta_2}^{\alpha_1\alpha_2}{e^{i\sum_{j=1}^2(\epsilon_{\alpha_j}-\epsilon_{\beta_j})t}}\!,
\end{equation}
so
\begin{equation}
\begin{split}
\label{eqO2}
    \langle\hat{\mathcal{O}}(2)\rangle_t^2 = \sum\limits_{\substack{\alpha_1,...,\alpha_4 \\ \beta_1,...,\beta_4}} & \tilde{\mathcal{O}}(1)^{(1)}_{\alpha_1\beta_1}\tilde{\mathcal{O}}(1)^{(2)}_{\alpha_2\beta_2} \tilde{\mathcal{O}}(1)^{(1)}_{\alpha_3\beta_3}\tilde{\mathcal{O}}(1)^{(2)}_{\alpha_4\beta_4}\\
    & \times R(2)_{\beta_1\beta_2}^{\alpha_1\alpha_2}R(2)_{\beta_3\beta_4}^{\alpha_3\alpha_4}{e^{i\sum_{j=1}^4(\epsilon_{\alpha_j}-\epsilon_{\beta_j})t}}.
\end{split}
\end{equation}
To compute the infinite-time average of Eq.~(\ref{eqO2}), we note that 
\begin{equation}
    \overline{e^{i\sum_{j=1}^2(\epsilon_{\alpha_j}-\epsilon_{\beta_j})t}}=\sum_{\sigma\in S_4}\prod_{j=1}^{4}\delta(\beta_j-\alpha_{\sigma(j)}),
\end{equation}
so that
\begin{equation}
\begin{split}
\label{eqO2_2}
    \overline{\langle\hat{\mathcal{O}}(2)\rangle_t^2} =& \sum\limits_{\substack{\sigma\in S_4,\\\alpha_1,...,\alpha_4}} \tilde{\mathcal{O}}(1)^{(1)}_{\alpha_1\alpha_{\sigma(1)}}\tilde{\mathcal{O}}(1)^{(2)}_{\alpha_2\alpha_{\sigma(2)}}\tilde{\mathcal{O}}(1)^{(1)}_{\alpha_3\alpha_{\sigma(3)}}\\
    & \times \tilde{\mathcal{O}}(1)^{(2)}_{\alpha_4\alpha_{\sigma(4)}} R(2)_{\alpha_{\sigma(1)}\alpha_{\sigma(2)}}^{\alpha_1\alpha_2} R(2)_{\alpha_{\sigma(3)}\alpha_{\sigma(4)}}^{\alpha_3\alpha_4},
\end{split}
\end{equation}
where $S_n$ is a permutation group of $n$ elements, while $\delta(\beta_i-\alpha_j)$ is a Kronecker delta. There are $4!=24$ contributions to the sum over $\sigma$ in Eq.~(\ref{eqO2_2}), which can be divided into the following classes:
\begin{enumerate}[label=(\Alph*), wide, labelwidth=!, labelindent=0pt]
    
    \item Both $\sigma(1)$ and $\sigma(2)$ belong to $\{1,2\}$. There are $4$ such contributions. Note that all these contributions appear in $\overline{\langle\hat{\mathcal{O}}(2)\rangle_t}^2$ and, thus, they are cancelled in the standard deviation.

    \item Either $\sigma(1)$ or $\sigma(2)$ belong to $\{1,2\}$. There are $16$ such contributions.

    \item Both $\sigma(1)$ and $\sigma(2)$ belong to $\{3,4\}$. There are $4$ such contributions.
    
\end{enumerate}

Next, we consider exemplary cases from classes (B) and (C). There are no significant changes in the calculations for different contributions from the same class.

We first consider the class (B) case: $\sigma(1)=1$, $\sigma(2)=3$, $\sigma(3)=2$ and $\sigma(4)=4$, whose upper bound is
\begin{equation}
   \text{max}_{\alpha_1,\alpha_2,i}\left\{\left|\tilde{\mathcal{O}}(1)^{(i)}_{\alpha_1\alpha_2}\right|^4\right\}\sum\limits_{\alpha_1,...,\alpha_4} R(2)_{\alpha_1\alpha_3}^{\alpha_1\alpha_2}R(2)_{\alpha_2\alpha_4}^{\alpha_3\alpha_4}.
\end{equation}
We can carry out partial traces $\sum_{\alpha_1}R(2)_{\alpha_1\alpha_3}^{\alpha_1\alpha_2}=\textbf{O}(V)R(1)_{\alpha_3}^{\alpha_2}$ and $\sum_{\alpha_4}R(2)_{\alpha_2\alpha_4}^{\alpha_3\alpha_4}=\textbf{O}(V)R(1)_{\alpha_2}^{\alpha_3}$ (see property 2). Therefore, the upper bound can be simplified to
\begin{equation}
\begin{split}
    & \text{max}_{\alpha_1,\alpha_2,i}\left\{\left|\tilde{\mathcal{O}}(1)^{(i)}_{\alpha_1\alpha_2}\right|^4\right\}\textbf{O}(V^2)\sum\limits_{\alpha_2,\alpha_3} R(1)_{\alpha_3}^{\alpha_2}R(1)_{\alpha_2}^{\alpha_3}=\\
    & \text{max}_{\alpha_1,\alpha_2,i}\left\{\left|\tilde{\mathcal{O}}(1)^{(i)}_{\alpha_1\alpha_2}\right|^4\right\}\textbf{O}(V^2)\sum\limits_{\alpha_2}
    [R(1)^2]_{\alpha_2}^{\alpha_2}=\\
    & \text{max}_{\alpha_1,\alpha_2,i}\left\{\left|\tilde{\mathcal{O}}(1)^{(i)}_{\alpha_1\alpha_2}\right|^4\right\}\textbf{O}(V^2) {\rm Tr} [R(1)^2].
\end{split}
\end{equation}
According to property 3, ${\rm Tr}[R(1)^2]\le\textbf{O}(V)$. With the help of assumption 1, we arrive at $\text{max}_{\alpha_1,\alpha_2,i}\{|\tilde{\mathcal{O}}(1)^{(i)}_{\alpha_1\alpha_2}|^4\}=\textbf{O}(V^{-4})$. Finally, the upper bound for the contribution from the class (B) is of order $\textbf{O}(V^{-4})\textbf{O}(V^2)\textbf{O}(V)=\textbf{O}(V^{-1})$.

Next, we consider the class (C) case: $\sigma(1)=3$, $\sigma(2)=4$, $\sigma(3)=1$, and $\sigma(4)=2$, whose upper bound is
\begin{equation}
\begin{split}
   & \text{max}_{\alpha_1,\alpha_2,i}\left\{\left|\tilde{\mathcal{O}}(1)^{(i)}_{\alpha_1\alpha_2}\right|^4\right\}\sum\limits_{\alpha_1,...,\alpha_4} R(2)_{\alpha_3\alpha_4}^{\alpha_1\alpha_2}R(2)_{\alpha_1\alpha_2}^{\alpha_3\alpha_4} = \\
   & \text{max}_{\alpha_1,\alpha_2,i}\left\{\left|\tilde{\mathcal{O}}(1)^{(i)}_{\alpha_1\alpha_2}\right|^4\right\} \sum_{\alpha_1\alpha_2} [R(2)^2]_{\alpha_1\alpha_2}^{\alpha_1\alpha_2}=\\
   & \text{max}_{\alpha_1,\alpha_2,i}\left\{\left|\tilde{\mathcal{O}}(1)^{(i)}_{\alpha_1\alpha_2}\right|^4\right\}{\rm Tr}[R(2)^2].
\end{split}
\end{equation}
The same properties and assumptions used for the case from the class (B), 3 and 1, allow us to write ${\rm Tr}[R(2)^2]\le\textbf{O}(V^3)$ and $\text{max}_{\alpha_1\alpha_2,i}\left\{\left|\tilde{\mathcal{O}}(1)^{(i)}_{\alpha_1\alpha_2}\right|^4\right\}=\textbf{O}(V^{-4})$. Consequently, the upper bound for the contribution from the class (C) is of order $\textbf{O}(V^{-4})\textbf{O}(V^3)=\textbf{O}(V^{-1})$.

We can now establish the standard deviation
\begin{equation}
    \sigma_t\left(\hat{O}(2)\right) \le \sigma_t\left(\hat{\mathcal{O}}(1)\right)+\sigma_t\left(\hat{\mathcal{O}}(2)\right)\le \textbf{O}(V^{-1/2}).
\end{equation}

\subsection{General case of $q$-body observables} \label{sec:q-body}

As shown in the previous section, we can express a $q$-body observable $\hat{O}(q)$ with a finite $q$ as a sum of $k$-body observables $\hat{\mathcal{O}}(k)^{(j)}$ with $k\in\left\{1,...,q\right\}$
\begin{equation}
    \hat{O}(q)=\sum_{j} \hat{\mathcal{O}}\left(k\right)^{(j)}
\end{equation}
where 
\begin{equation}
    \hat{\mathcal{O}}(k)^{(j)}\!=\hspace{-0.2cm}\sum\limits_{\substack{\alpha_1,...,\alpha_k \\ \beta_1,...,\beta_k}}\! \tilde{\mathcal{O}}(1)^{(j,1)}_{\alpha_1\beta_1}...\;\tilde{\mathcal{O}}(1)^{(j,k)}_{\alpha_k\beta_k} \hat{f}_{\alpha_1}^\dagger...\; \hat{f}_{\alpha_k}^\dagger\hat{f}_{\beta_k}...\;\hat{f}_{\beta_1}.
\end{equation} 
Note that for $q\ge 3$, $k$ is generally not equal to $j$. For example, for $q=3$, there is exactly one one-body observable $\hat{\mathcal{O}}(1)^{(1)}$, three two-body observables $\hat{\mathcal{O}}(2)^{(2)}$, $\hat{\mathcal{O}}(2)^{(3)}$, and $\hat{\mathcal{O}}(2)^{(4)}$, and one three-body observable $\hat{\mathcal{O}}(3)^{(5)}$.

Thanks to property 4, we can focus on a standard deviation of a single operator $\hat{\mathcal{O}}(k)^{(j)}$
\begin{equation}
    \sigma_t(\hat{\mathcal{O}}(k))=\sqrt{\overline{\langle\hat{\mathcal{O}}(k)\rangle_t^2}-\overline{\langle\hat{\mathcal{O}}(k)\rangle}_t^2}.
\end{equation}
As before, we drop the index $j$ in the observable $\hat{\mathcal{O}}(k)^{(j)}\rightarrow \hat{\mathcal{O}}(k)$ and matrices $\tilde{\mathcal{O}}_{\alpha_1,\beta_1}^{(j,i)}\rightarrow \tilde{\mathcal{O}}_{\alpha_1,\beta_1}^{(i)}$, and we use a simplified notation $\langle\hat{\mathcal{O}}(k)^{(j)}(t)\rangle\rightarrow\langle\hat{\mathcal{O}}(k)\rangle_t$.

The expectation value of $\hat{\mathcal{O}}(k)$ at time $t$ is given by
\begin{equation}
\begin{split}
    \langle\hat{\mathcal{O}}(k)\rangle_t=\sum\limits_{\substack{\alpha_1,...,\alpha_k \\ \beta_1,...,\beta_k}} & \tilde{\mathcal{O}}(1)^{(1)}_{\alpha_1\beta_1}...\;\tilde{\mathcal{O}}(1)^{(k)}_{\alpha_k\beta_k} R(k)_{\beta_1...\beta_k}^{\alpha_1...\alpha_k}\\
    & \times {e^{i\sum_{j=1}^k(\epsilon_{\alpha_j}-\epsilon_{\beta_j})t}},
\end{split}
\end{equation}
so
\begin{equation}
\begin{split}
\label{eqO2_3}
    &\langle\hat{\mathcal{O}}(k)\rangle_t^2 = \hspace{-0.2cm}\sum\limits_{\substack{\alpha_1,...,\alpha_{2k} \\ \beta_1,...,\beta_{2k}}} \hspace{-0.1cm} \tilde{\mathcal{O}}(1)^{(1)}_{\alpha_1\beta_1}...\;\tilde{\mathcal{O}}(1)^{(k)}_{\alpha_{k}\beta_{k}}\tilde{\mathcal{O}}(1)^{(1)}_{\alpha_{k+1}\beta_{k+1}}...\\ &\times \hspace{-0.1cm} \;\tilde{\mathcal{O}}(1)^{(k)}_{\alpha_{2k}\beta_{2k}}
    R(k)_{\beta_1...\beta_{k}}^{\alpha_1...\alpha_{k}}R(k)_{\beta_{k+1}...\beta_{2k}}^{\alpha_{k+1}...\alpha_{2k}}
     {e^{i\sum_{j=1}^{2k}(\epsilon_{\alpha_j}-\epsilon_{\beta_j})t}},
\end{split}
\end{equation}
whose infinite-time average is
\begin{equation}
\begin{split}
    \label{eqO2_4}
    &\overline{\langle\hat{\mathcal{O}}(k)\rangle_t^2} \! = \hspace{-0.4cm}\sum\limits_{\substack{\sigma\in S_{2k}\\ \alpha_1,...,\alpha_{2k}}} \hspace{-0.3cm} \tilde{\mathcal{O}}(1)^{(1)}_{\alpha_1\alpha_{\sigma(1)}}...\;\tilde{\mathcal{O}}(1)^{(k)}_{\alpha_{k}\alpha_{\sigma(k)}}
    \tilde{\mathcal{O}}(1)^{(1)}_{\alpha_{k+1}\alpha_{\sigma(k+1)}}...\;\\& \times\tilde{\mathcal{O}}(1)^{(k)}_{\alpha_{2k}\alpha_{\sigma(2k)}}
     R(k)_{\alpha_{\sigma(1)}...\alpha_{\sigma(k)}}^{\alpha_1...\alpha_{k}} R(k)_{\alpha_{\sigma(k+1)}...\alpha_{\sigma(2k)}}^{\alpha_{k+1}...\alpha_{2k}}.
\end{split}
\end{equation}
There are $(2k)!$ contributions to the sum over $\sigma$ in Eq.~(\ref{eqO2_4}). They can be divided into $k+1$ classes. In each class, there are $r\in\{0,...,k\}$ pairs of identical indices $\alpha_i=\alpha_{\sigma(j)}$ with $i,j\in\{1,...,k\}$. The class with $r=k$ is cancelled in the standard deviation. For the class with $r<k$, carrying out all the partial traces of $k$-body density matrices over the repeated indices, results in the following upper bound
\begin{equation}
    \text{max}_{\alpha_1,\alpha_2,i}\left\{\left|\hat{\mathcal{O}}(1)_{\alpha_1\alpha_2}^{(i)}\right|^{2k}\right\}\textbf{O}(V^{2r})\text{Tr}\left[R(k-r)^2\right].
\end{equation}
According the property 3,
\begin{equation}
    \text{Tr}\left[R(k-r)^2\right]\le
    \left\{
    \begin{array}{ll}
    \textbf{O}(V^{3(k-r)/2}) & \text{ for }k-r = \text{even},\\ 
    \textbf{O}(V^{3(k-r)/2-1/2}) & \text{ for }k-r = \text{odd}.
    \end{array}
    \right.
\end{equation}
It is clear that only classes with $r=k-1$ and $r=k-2$ can be of the highest order. In the former case
\begin{equation}
\begin{split}
    & \text{max}_{\alpha_1,\alpha_2,i}\left\{\left|\hat{\mathcal{O}}(1)_{\alpha_1\alpha_2}^{(i)}\right|^{2k}\right\}\textbf{O}(V^{2(k-1)})\text{Tr}\left[R(1)^2\right]\le\\
    & \textbf{O}\left(V^{-2k}\right) \textbf{O}\left(V^{2k-2}\right) \textbf{O}\left(V\right)=\textbf{O}\left(V^{-1}\right),
\end{split}
\end{equation}
while in the latter case
\begin{equation}
\begin{split}
    & \text{max}_{\alpha_1,\alpha_2,i}\left\{\left|\hat{\mathcal{O}}(1)_{\alpha_1\alpha_2}^{(i)}\right|^{2k}\right\}\textbf{O}(V^{2(k-2)})\text{Tr}\left[R(2)^2\right]\le\\
    & \textbf{O}\left(V^{-2k}\right) \textbf{O}\left(V^{2k-4}\right) \textbf{O}\left(V^3\right)=\textbf{O}\left(V^{-1}\right).
\end{split}
\end{equation}
Therefore, the standard deviation is upper bounded by
\begin{equation}
    \sigma_t\left(\hat{O}(q)\right)\le\sum_{j}\sigma_t\left(\hat{\mathcal{O}}(k)^{(j)}\right)\le\textbf{O}\left(V^{-1/2}\right),
\end{equation}
where we have reintroduced the upper index for observables $\hat{\mathcal{O}}(k)^{(j)}$. We have, thus, proved that equilibration is guaranteed in quantum-chaotic quadratic models not only for one-body but also for multi-particle observables. It is remarkable that the established order of the upper bound for $\sigma_t(\hat{O}(q))$ is independent of $q$.

\section{Absence of eigenstate thermalization in many-body energy eigenstates}

The diagonal matrix elements of the one-body observable $\hat{O}$ in the many-body energy eigenstates $|\Omega\rangle$ can be written as
\begin{equation}
\label{eq_diag_mb}
    \langle\Omega|\hat{O}|\Omega\rangle = \sum_{\alpha,\beta=1}^{V} O_{\alpha\beta}\langle\Omega|\hat{f}^\dagger_{\alpha}\hat{f}^{}_{\beta}|\Omega\rangle=\sum_{\alpha=1}^{V} O_{\alpha\alpha}\langle\Omega|\hat{f}^\dagger_{\alpha}\hat{f}^{}_{\alpha}|\Omega\rangle\;.
\end{equation}
As shown for QCQ Hamiltonians in Ref.~\cite{lydzba_zhang_21}, the diagonal matrix elements of $\hat{O}$ in the single-particle energy eigenstates $|\alpha\rangle$ are described by the ETH ansatz
\begin{equation}
\label{eq_diag_sp}
    \frac{O_{\alpha\alpha}}{||\hat{O}||}=\mathcal{O}\left(\epsilon_\alpha\right)+\rho\left(\epsilon_\alpha\right)^{-1/2} \mathcal{F}_{O}\left(\epsilon_\alpha,0\right) R^{O}_{\alpha\alpha}\;,
\end{equation}
where $||\hat{O}||^2 = \frac{1}{V} \sum_{\alpha,\beta=1}^{V} O_{\alpha\beta}^2$ is the Hilbert-Schmidt norm in the single-particle Hilbert space, $\mathcal{O} \left(\epsilon_\alpha\right)$ and $\mathcal{F}_{O} \left(\epsilon_\alpha,0\right)$ are smooth functions of single-particle energy, and $R^O_{\alpha\alpha}$ is a random number with zero mean and unit variance. For one-body observables whose rank is ${\bf O}(1)$, our focus here, $||\hat{O}||^2 \propto \frac{1}{V}$. For those observables is convenient to rewrite Eq.~(\ref{eq_diag_sp}) as
\begin{equation}\label{eq:eth1resc}
    O_{\alpha\alpha}=\frac{O\left(\epsilon_\alpha\right)}{\sqrt{V}}+ \frac{{F}_{O}\left(\epsilon_\alpha\right)}{V} R^{O}_{\alpha\alpha}\;,
\end{equation}
where we have used $\rho(\epsilon_\alpha)\propto V$, and we have introduced the rescaled functions $O\left(\epsilon_\alpha\right) = \sqrt{V} ||\hat{O}|| \mathcal{O} \left(\epsilon_\alpha\right)$ and ${F}_{O}\left(\epsilon_\alpha\right) = V ||\hat{O}|| \rho \left(\epsilon_\alpha\right)^{-1/2} \mathcal{F}_{O} \left(\epsilon_\alpha,0\right)$. This notation makes the scaling with the system size explicit. Finally, using Eq.~\eqref{eq:eth1resc}, we can rewrite Eq.~(\ref{eq_diag_mb}) as
\begin{equation}
\begin{split}
\label{eq_terms}
     \langle\Omega|\hat{O}|\Omega\rangle & = \sum_{\alpha=1}^{V}\frac{O\left(\epsilon_\alpha\right)}{\sqrt{V}} \langle\Omega|\hat{f}^\dagger_{\alpha}\hat{f}_{\alpha}|\Omega\rangle\\
     & +\sum_{\alpha=1}^{V}\frac{{F}_{O}\left(\epsilon_\alpha\right)}{V} R^{O}_{\alpha\alpha} \langle\Omega|\hat{f}^\dagger_{\alpha}\hat{f}_{\alpha}|\Omega\rangle\;.
\end{split}
\end{equation}

We now focus on the second term on the r.h.s.~of Eq.~(\ref{eq_terms}), which is a sum of $N$ independent random numbers with zero mean $\mu_\alpha=0$ and variance $\sigma^2_\alpha = \left( F_O \left(\epsilon_\alpha\right)/V \right)^2$. It satisfies Lindeberg's condition 
\begin{equation}
    \lim_{N\rightarrow\infty}
    \sum_{\beta=1}^{N}
    \frac{\mathbb{E}\left[\left(\frac{{F}_{O}\left(\epsilon_\beta\right)}{V} R^{O}_{\beta\beta}\right)^2,\left|\frac{{F}_{O}\left(\epsilon_\beta\right)}{V} R^{O}_{\beta\beta}\right|>\epsilon\sigma_N\right]}{\sigma_N^2}=0,
\end{equation}
where $\epsilon$ is an arbitrary positive number, $\mathbb{E}\left(...\right)$ stands for the expectation value, while $\beta$ runs over the single-particle energy eigenstates with $\langle\Omega|\hat{f}^\dagger_{\beta}\hat{f}^{}_{\beta}|\Omega\rangle=1$. Additionally, the total mean is $\mu_N=\sum_{\beta=1}^{N}\mu_\beta=0$ and the total variance is
\begin{equation}
    \sigma^2_N=\sum_{\beta=1}^{N}\sigma^2_\beta=\sum_{\beta=1}^{N} \left[{F}_{O}\left(\epsilon_\beta\right)/V\right]^2\propto N/V^2 = \bar{n}/V\;.
\end{equation} 
Note that Lindeberg's condition is a sufficient condition for the central limit theorem to hold for a sequence of independent random variables, and it is equivalent to the requirement that none of these random variables has a variance $\sigma^2_\beta$ that is a non-vanishing fraction of the total variance $\sigma^2_N$. Therefore, the second term in the sum in Eq.~(\ref{eq_terms}) is a random number from a normal distribution with $\mu_N=0$ and $\sigma_N^2\propto \bar{n}/V$. Consequently, the diagonal matrix elements of one-body observables $\hat{O}$ in the many-body energy eigenstates $|\Omega\rangle$ exhibit fluctuations that vanish at most polynomially with increasing the system size. This statement is consistent with the analysis in the SYK2 model~\cite{haque_mcclarty_19}.

More importantly, there is an exponentially large number of outliers, i.e., $\langle\Omega|\hat{O}|\Omega\rangle$ that exhibit a nonvanishing difference with the microcanonical average in the thermodynamic limit. Again, we focus on the second term on the r.h.s.~of Eq.~(\ref{eq_terms}).  We assume that for $\alpha$'s for which $\langle\Omega| \hat{f}^\dagger_{\alpha} \hat{f}^{}_{\alpha} |\Omega\rangle=1$, there are $\frac{V}{a}$ $\left(\frac{V}{b}\right)$ positive (negative) values of $\frac{{F}_{O}\left(\epsilon_\alpha\right)}{V} R^{O}_{\alpha\alpha}$, where $a$ and $b$ are ${\bf O}(1)$ and $\frac{1}{a}+\frac{1}{b}=\bar{n}$. We can now divide the second term in Eq.~(\ref{eq_terms}) in two parts,
\begin{equation}
    \langle\Omega|\hat{O}|\Omega\rangle\propto \sum_{\beta=1}^{V/a}\left|\frac{{F}_{O}\left(\epsilon_\beta\right)}{V} R^{O}_{\beta\beta}\right|
    -
   \sum_{\gamma=1}^{V/b}\left|\frac{{F}_{O}\left(\epsilon_\gamma\right)}{V} R^{O}_{\gamma\gamma}\right|\;.
\end{equation}
The first (second) term in the previous equation is a sum of $\frac{V}{a}$ $\left(\frac{V}{b}\right)$ independent random numbers with mean $\mu_\beta\propto\frac{1}{V}$ $\left(\mu_\gamma\propto\frac{1}{V}\right)$ and variance $\sigma^2_\beta\propto\frac{1}{V^2}$ $\left(\sigma^2_\gamma\propto\frac{1}{V^2}\right)$. Therefore, it is a random number from a normal distribution with mean $\mu_{V/a}\propto\frac{1}{a}$ $\left(\mu_{V/b}\propto\frac{1}{b}\right)$ and variance $\sigma^2_{V/a}\propto\frac{1}{aV}$ $\left(\sigma^2_{V/b}\propto\frac{1}{bV}\right)$. Consequently, the considered $\langle\Omega|\hat{O}|\Omega\rangle$ have the following expectation value:
\begin{equation}
    \mathbb{E}\left(\langle\Omega|\hat{O}|\Omega\rangle)\right)\propto \frac{1}{a}-\frac{1}{b}\;,
\end{equation}
while the variance is independent of $a$ and $b$ and, so, the same as in the previous paragraph. We arrive at the conclusion that the majority of $\langle\Omega|\hat{O}|\Omega\rangle$ for an arbitrary finite difference $\frac{1}{a}-\frac{1}{b}\neq 0$ are outliers. 

We can use a simple combinatorial argument to estimate the lower bound for the number of outliers $\mathcal{N}$. Let us assume that half of $\frac{{F}_{O}\left(\epsilon_\alpha\right)}{V} R^{O}_{\alpha\alpha}$ are positive and half are negative. For fixed $a$ and $b$,
\begin{equation}
\begin{split}
    \mathcal{N} & =
    \begin{pmatrix}
    V/2\\ 
    V/a
    \end{pmatrix}
    \begin{pmatrix}
    V/2\\ 
    V/b
    \end{pmatrix}
    \ge
    \left[\frac{V/2}{V/a}\right]^\frac{V}{a} \left[\frac{V/2}{V/b}\right]^\frac{V}{b}\\
    & = \left[\left(\frac{a}{2}\right)^{\frac{1}{a}}\right]^{V} \left[\left(\frac{b}{2}\right)^{\frac{1}{b}}\right]^{V}
    =2^{\kappa V}\;,
\end{split}
\end{equation}
where we have introduced $\kappa =\frac{1}{a}\log_2\left(\frac{a}{2}\right)+\frac{1}{b}\log_2\left(\frac{b}{2}\right)$. Hence, the number $\mathcal{N}$ is exponentially large for the physically relevant parameters $a, b\ge 2$ (because $\frac{1}{a}+\frac{1}{b}=\bar{n}$), recalling that we need $a\neq b$ so that $\frac{1}{a}-\frac{1}{b}\neq 0$.

\section{SYK2 model}

The SYK2 model is usually written in the form
\begin{equation}
\label{def_syk2}
    \hat{H}=\sum_{i,j=1}^{V} a_{ij} \hat{c}_{i}^\dagger \hat{c}^{}_{j},
\end{equation}
where the diagonal (off-diagonal) elements of the matrix ${\bf a}$ are real normally distributed random numbers with zero mean and $2/V$ ($1/V$) variance. A quench to the SYK2 model from Eq.~(\ref{def_syk2}) is a strong quench to the ``infinite-temperature'' regime, independently of the initial state chosen.

In order to control the strength of the quench in the context of this model, we have modified it's Hamiltonian
\begin{equation}\label{eq:swk2m}
    \hat{H}=\sum_{i,j=1}^{V} \left[(1-\gamma)a_{ij}+\gamma b_{ij}\right] \hat{c}_{i}^\dagger \hat{c}_{j},
\end{equation}
where the diagonal (off-diagonal) matrix elements of both ${\bf a}$ and ${\bf b}$ are real normally distributed random numbers with zero mean and $2/V$ ($1/V$) variance, while $\gamma\in[0,1]$. We then only quench the matrix ${\bf b}$ to a different realization, so that if $\gamma\ll 1$ the quench is weak and if $\gamma\rightarrow1$ the quench is strong.

\begin{figure}[!t]
\includegraphics[width=\columnwidth]{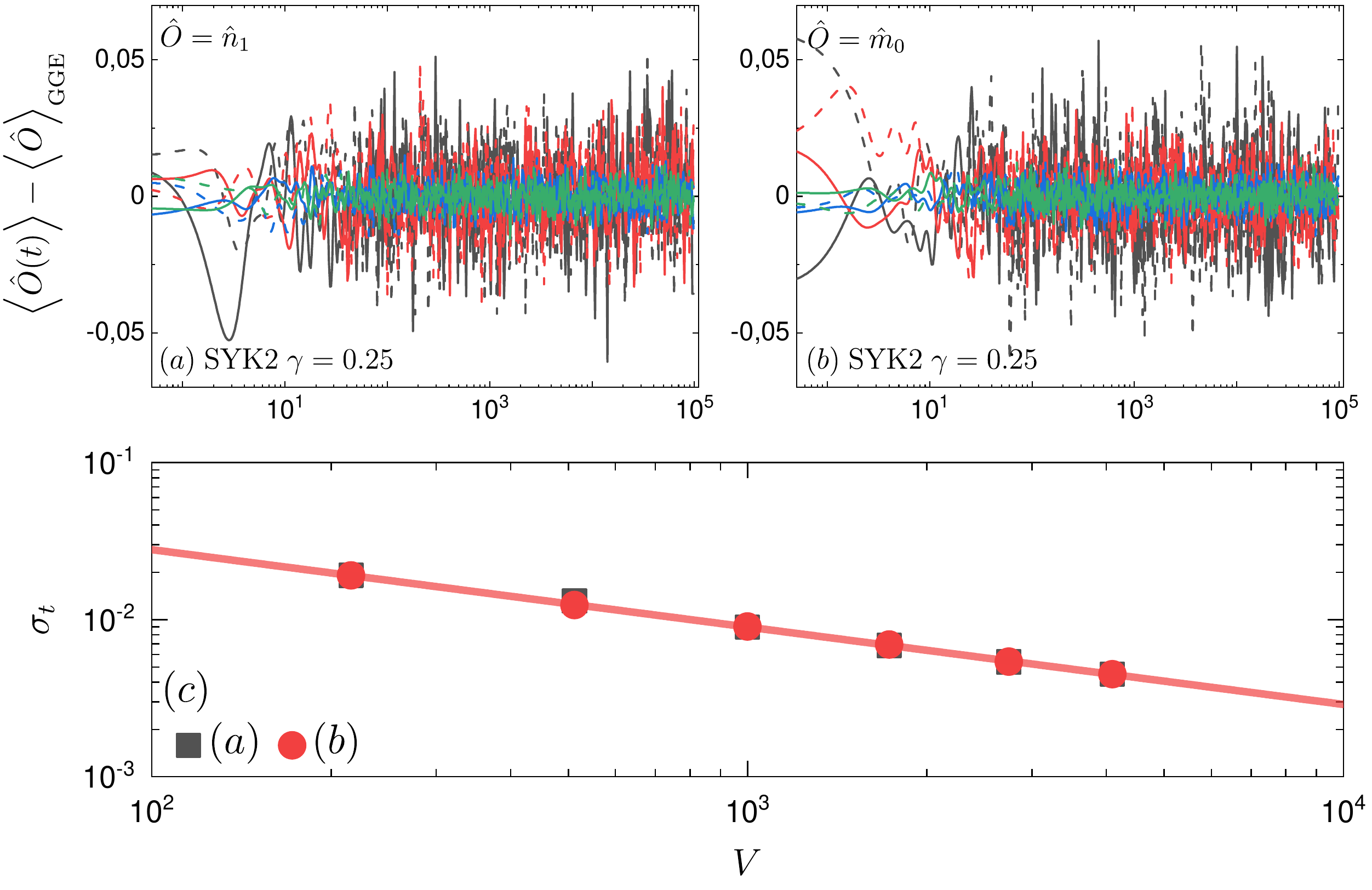}
\vspace{-0.5cm}
\caption{(a),(b) Time evolution of $\langle\hat{O}(t)\rangle - \langle\hat{O}\rangle_\text{GGE}$ for a quench within the SYK2 model with $\gamma=0.25$ and $\bar{n}=0.5$. The numerical results for system with $V=6^3,\, 8^3,\, 14^3,$ and $16^3$ are marked with black, red, blue, and green lines, respectively. We show results for two (solid and dashed) quench realizations for each $V$. Two operators are considered (a) $\hat n_1$ and (b) $\hat m_0$. (c) Temporal fluctuations $\sigma_t$ calculated within the time interval $t\in[10^2, 10^5]$ and averaged over $20$ quench realizations. The lines show the outcome of two parameter fits $\kappa/V^\zeta$. We get $\zeta\in[0.49,0.5]$ for (a) and (b).}
\label{figS0}
\end{figure}

In Figs.~\ref{figS0}(a) and~\ref{figS0}(b), we show the time evolution of $\hat{n}_{1}$ and of $\hat{m}_{0}$ in a quench within the SYK2 model. We consider $\gamma=0.25$ and $\bar{n}=0.5$. Note that the temporal fluctuations $\sigma_t$ decrease with increasing the number of lattice sites $V$. As confirmed in Fig.~\ref{figS0}(a), $\sigma_t\propto V^{-\zeta}$ with $\zeta\approx0.5$. The exponent $\zeta=0.5$ is expected, since the SYK2 model is a paradigmatic example of a quadratic quantum-chaotic model.

\begin{figure}[!b]
\includegraphics[width=0.98\columnwidth]{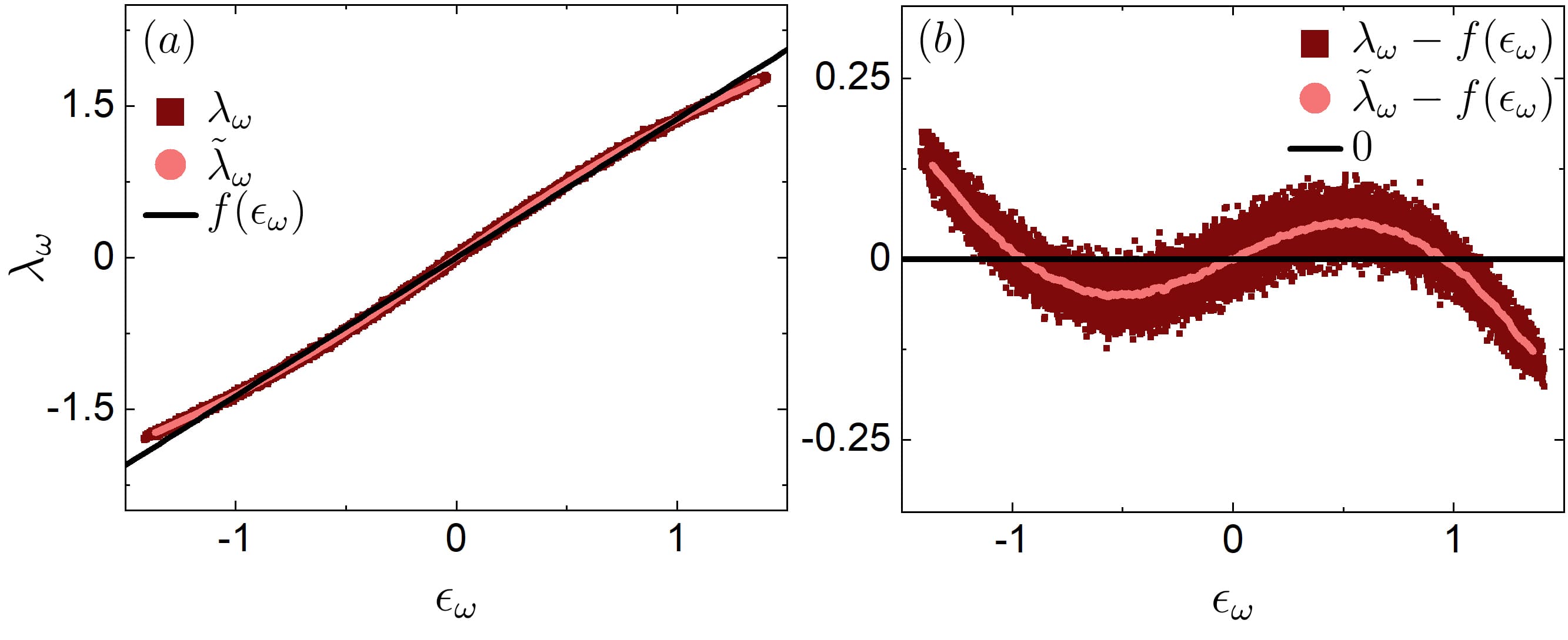}
\vspace{-0.2cm}
\caption{(a) Lagrange multipliers $\lambda_\omega$ as functions of single-particle energies $\epsilon_\omega$ for a quench within the SYK2 model with $\gamma=0.5$ and $\bar{n}=1/2$. Dark red squares mark results for a single quench realization, which were previously reported in the inset of Fig.~\ref{fig3}(a). We also show (light red circles) the moving average $\tilde{\lambda}_\omega$, and (straight black line) the result of a least-squares fit $f(\epsilon_\omega)=1.37 \epsilon_\omega$. (b) Differences between the Lagrange multipliers $\lambda_\omega$ and the fit $f(\epsilon_\omega)$.}
\label{figS1}
\end{figure}

As discussed in the main text, the infinite-time averages of one-body observables in quadratic models are described by the GGE. When the Lagrange multipliers~$\lambda_\alpha$ are linear in the single-particle energies~$\epsilon_\alpha$, the GGE and the GE are identical. For our quenches within the SYK2 model from Eq.~\eqref{eq:swk2m}, we find that the differences between the GGE and the GE decrease as $\gamma\rightarrow1$, but the two ensembles remain different unless $\gamma = 1$. A similar behavior (GGEs that approach GEs) has been observed in the context of other families of quenches in integrable models mappable to noninteracting ones~\cite{Fitz_2011, He_2012, He_2013b}.

In Fig.~\ref{figS1}(a), the dark red squares show the Lagrange multipliers $\lambda_\alpha$ previously reported in the inset of Fig.~\ref{fig3}(a) for $\gamma=0.5$ ($V=28^3$), the light red circles show the moving average $\tilde{\lambda}_{\alpha} = \sum_{\beta = \alpha - 99}^{\alpha + 100} \lambda_\beta / 200$, while the continuous black line shows the result of a least-squares fit to $f(\epsilon_\alpha) = \text{const}\times \epsilon_\alpha$. The deviations of the data from the fit are apparent. We plot them in Fig.~\ref{figS1}(b). With increasing system size, the eigenstate-to-eigenstate fluctuations of $\lambda_\alpha$ decrease and the Lagrange multipliers $\lambda_\alpha$ approach the moving average $\tilde{\lambda}_{\alpha}$, i.e., the deviations from the fit $f(\epsilon_\alpha)$ do not vanish. Qualitatively similar results were obtained for other values of $\gamma<1$.

\section{Smoothness of the Lagrange multipliers}

\begin{figure}[!b]
\includegraphics[width=0.98\columnwidth]{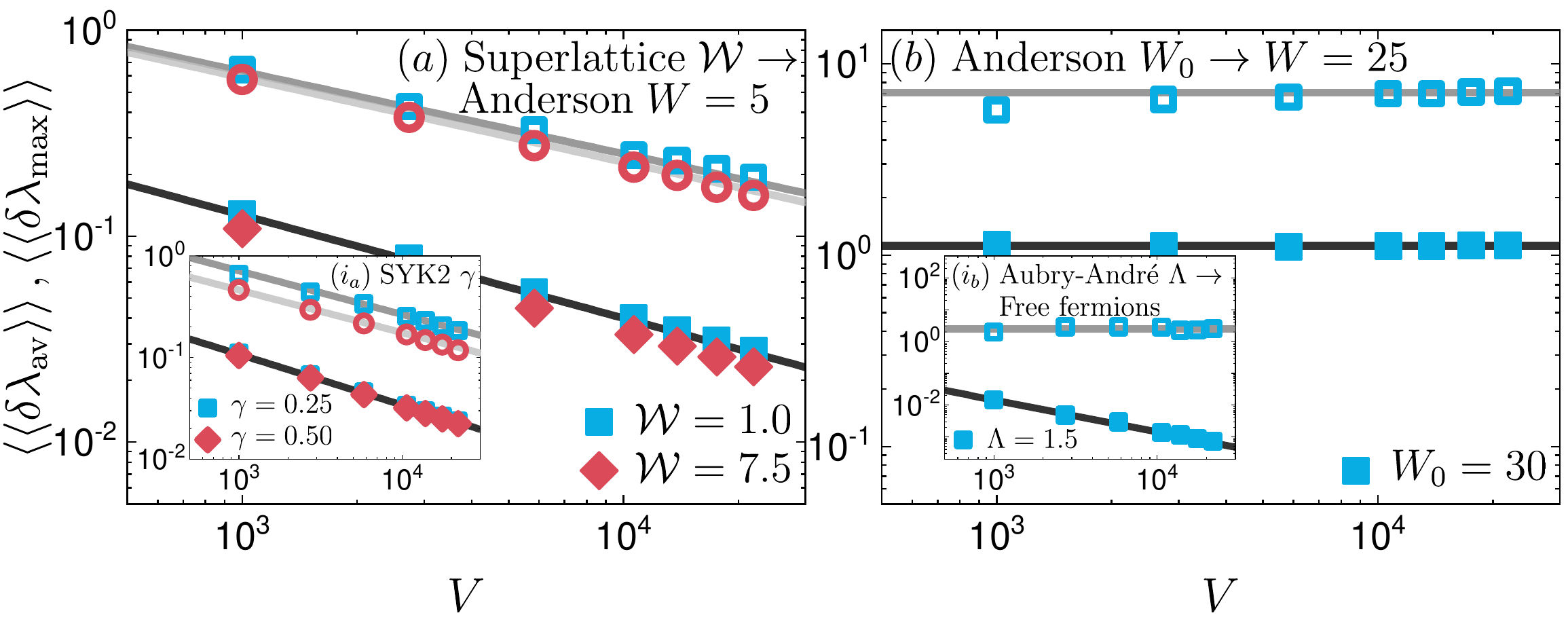}
\vspace{-0.2cm}
\caption{Eigenstate-to-eigenstate fluctuations of $\lambda_\alpha$. The considered quenches are: (a) the 3D superlattice model with ${\cal W}=1$ and 7.5 to the 3D Anderson model with $W=5$ (main panel), and the change of ${\bf b}$ to a new random realization in the SYK2 model with $\gamma=0.25$ and 0.5 (inset); (b) the 3D Anderson model at $W_0=30$ to the same model (with a different disorder realization) at $W=25$ (main panel), and the Aubry-Andr\'e model with $\Lambda=1.5$ to free fermions (inset). Open and closed symbols correspond to $\langle \langle \delta \lambda^\text{max} \rangle \rangle$ and $\langle \langle \delta \lambda^\text{av} \rangle \rangle$, respectively. Solid lines show the results of two parameter fits $\kappa/V^\zeta$. Whenever $\langle \langle \delta\lambda^\text{max} \rangle \rangle$ and $\langle \langle \delta\lambda^\text{av} \rangle \rangle$ decrease with $V$, we get $\zeta\in[0.4,0.5]$.}
\label{figS2}
\end{figure}

In this section, we study the eigenstate-to-eigenstate fluctuations of the Lagrange multipliers, $\delta\lambda_\alpha=\lambda_\alpha-\lambda_{\alpha-1}$. We consider both the average and the maximal value of $|\delta\lambda_\alpha|$,
\begin{equation} \label{def_e_to_e_fluctuations}
\delta\lambda^\text{av}=\frac{1}{V-1}\sum_{\alpha=2}^{V} |\delta\lambda_\alpha|,\;\;
\delta\lambda^\text{max}=\text{max}\left\{|\delta\lambda_\alpha|\right\} \;.
\end{equation}
In all calculations, we first determine $\delta \lambda^\text{av}$ and $\delta \lambda^\text{max}$ for a single quench and then average over $50$ quench realizations yielding $\langle \langle \delta\lambda^\text{av} \rangle \rangle$ and $\langle \langle \delta\lambda^\text{max} \rangle \rangle$, respectively. The finite size scaling of the eigenstate-to-eigenstate fluctuations for quenches in which the final Hamiltonian exhibits single-particle quantum chaos is presented in Fig.~\ref{figS2}(a): the 3D Anderson model with $W = 5$ (main panel) and the SYK2 model (inset). As expected from Fig.~\ref{fig3}, $\langle \langle \delta\lambda^\text{av} \rangle \rangle$ and $\langle \langle \delta\lambda^\text{max} \rangle \rangle$ decay as $\propto 1/V^\zeta$ with $\zeta\in[0.4,0.5]$. The finite-size scaling of the eigenstate-to-eigenstate fluctuations for quenches in which the final Hamiltonian does not exhibit single-particle quantum chaos is presented in Fig.~\ref{figS2}(b): the 3D Anderson model in the localized regime with W = 25 (main panel) and 1D noninteracting fermions in a homogeneous potential (inset). In this case, $\langle \langle \delta\lambda^\text{max} \rangle \rangle$ ($\langle \langle \delta\lambda^\text{av} \rangle \rangle$) does not (may or may not) decay with $V$.

\begin{figure}[!t]
\includegraphics[width=0.98\columnwidth]{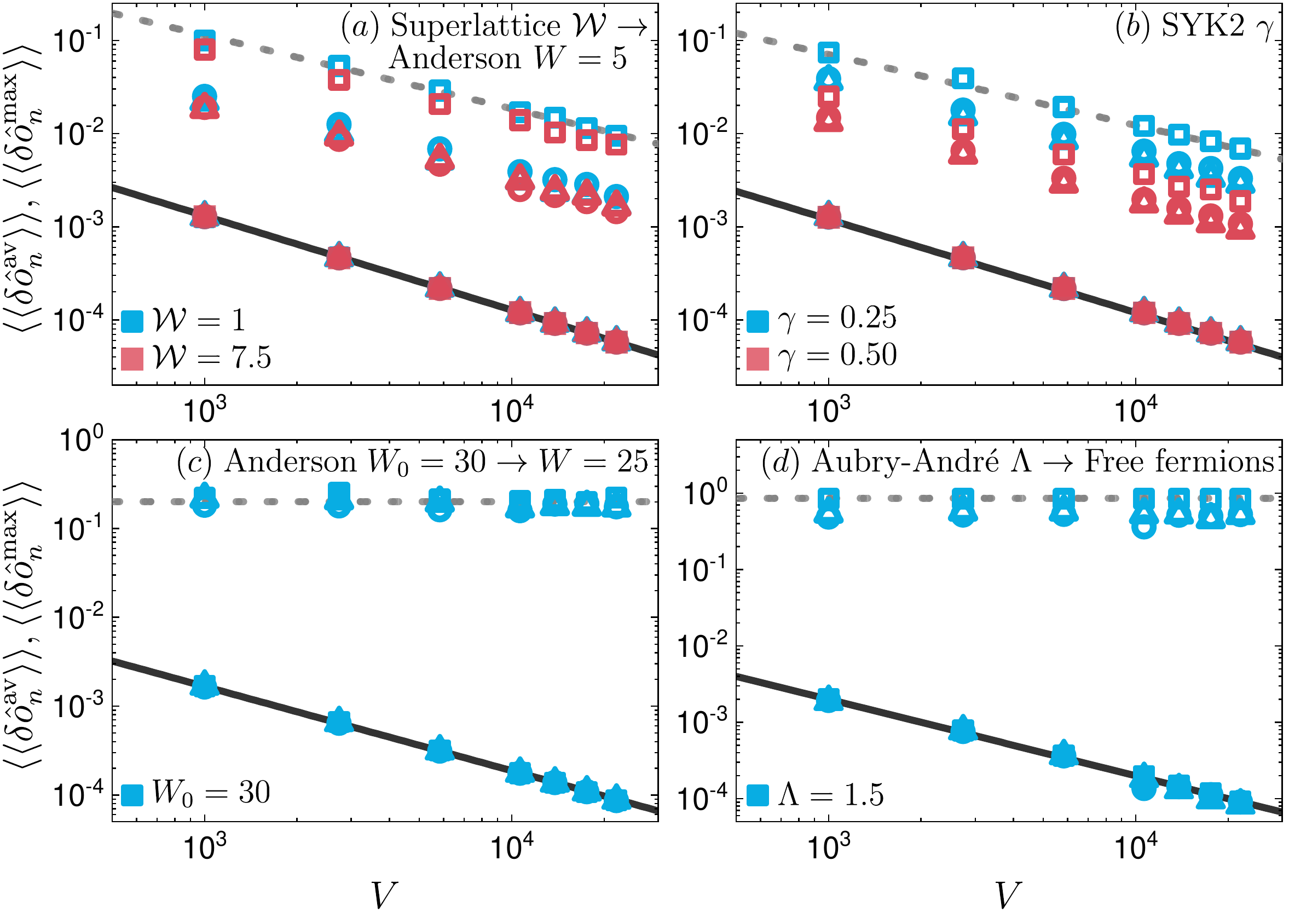}
\caption{Eigenstate-to-eigenstate fluctuations of $\hat o_n$ at $n=2,\, V/4$ and $V/2$ marked by squares, circles and triangles, respectively. We consider the same quenches as in Fig.~\ref{fig3} (see legends). Open and closed symbols correspond to $\langle \langle \delta \hat{o}_{n}^\text{max} \rangle \rangle$ and $\langle \langle \delta \hat{o}_{n}^\text{av} \rangle \rangle$, respectively. The latter are defined by replacing $\delta \lambda_\alpha$ with $\delta (o_n)_\alpha$ in Eq.~(\ref{def_e_to_e_fluctuations}). Dashed and solid lines are the two parameter fits of $\kappa/V^\zeta$ to $\langle \langle \delta \hat{o}_{n}^\text{max} \rangle \rangle$ and $\langle \langle \delta \hat{o}_{n}^\text{av} \rangle \rangle$ for $n=2$, respectively. If the fluctuations decrease with the system size $V$, they scale as $\propto 1/V^\zeta$ with $\zeta\in[0.6,1.0]$.}
\label{figS3}
\end{figure}

The smoothness of the Lagrange multipliers can be traced back to single-particle eigenstate thermalization. Recall that the Lagrange multipliers have values $\lambda_\alpha = \ln[(1 - \langle\hat{I}_\alpha\rangle) / \langle\hat{I}_\alpha\rangle]$~\cite{rigol_dunjko_07}. Hence, for $\delta\lambda_\alpha$ to vanish with increasing system size one needs $\delta \langle \hat{I}_\alpha \rangle = \langle\hat{I}_\alpha\rangle - \langle\hat{I}_{\alpha-1}\rangle$ to vanish as well. This can be seen from the Taylor expansion of the logarithm in $\delta \langle \hat{I}_\alpha \rangle$ near $\delta \langle \hat{I}_\alpha \rangle=0$.

Let us consider $|\Psi_0\rangle = \prod_{\{n\}} \hat a_n^\dagger |\emptyset\rangle$. For the quantum quenches studied in the previous sections, $|\Psi_0\rangle$ is the ground state of the initial Hamiltonian, while $\{n\}$ runs over its Fermi sea. Therefore, $\langle\hat{I}_\alpha\rangle = \sum_{\left\{n\right\}} \langle n | \alpha\rangle\langle\alpha |n\rangle$ can be expressed as $\langle\hat{I}_\alpha\rangle = \sum_{\left\{n\right\}} \langle\alpha | \hat{o}_{n} |\alpha\rangle$, where $\hat{o}_{n}=\hat{a}_n^\dagger \hat{a}_n = |n \rangle\langle n|$ are occupation operators of $|n\rangle$. Moreover, $\delta \langle \hat{I}_\alpha \rangle$ is governed by the eigenstate-to-eigenstate fluctuations, $\delta \langle \hat{I}_\alpha \rangle = \sum_{\left\{n\right\}} \delta (o_n)_{\alpha}$, where $(o_n)_{\alpha} = \langle\alpha | \hat{o}_{n} |\alpha\rangle$ are the diagonal single-particle matrix elements of $\hat o_n$, and $\delta (o_n)_{\alpha} = (o_n)_{\alpha} - (o_n)_{\alpha-1}$.

We note that the operators $\hat{o}_{n}$, being occupation operators, are expected to exhibit eigenstate thermalization in QCQ models as $\hat n_i$ and $\hat m_0$ do. Specifically, one expects the eigenstate-to-eigenstate fluctuations to decay as $1/V$. (The traceless normalized operator corresponding to $\hat o_n$ has the form $\underline{\hat{o}}_n=\frac{1}{\sqrt{V-1}}\left(V \hat o_n-1\right)$, for which $\delta(\underline{o}_n)_\alpha \propto 1/\sqrt{V}$~\cite{lydzba_zhang_21}.) Consequently, as a result of the central limit theorem, the standard deviation of the extensive sum $\delta \langle \hat{I}_\omega \rangle$ is expected to scale $\propto 1/\sqrt{V}$, as observed in Fig.~\ref{figS2}(a).

We confirm these predictions by numerical calculations reported in Fig.~\ref{figS3}. We consider the quantum quenches from Fig.~\ref{fig3}, and compute the eigenstate-to-eigenstate fluctuations $\langle \langle \delta \hat{o}_{n}^\text{av} \rangle \rangle$ and $\langle \langle \delta \hat{o}_{n}^\text{max} \rangle \rangle$ in analogy to Eq.~(\ref{def_e_to_e_fluctuations}). The eigenstate-to-eigenstate fluctuations in the presence of single-particle quantum chaos are plotted in Fig.~\ref{figS3}(a) and~\ref{figS3}(b). Both $\langle \langle \delta \hat{o}_{n}^\text{av} \rangle \rangle$ and $\langle \langle \delta \hat{o}_{n}^\text{max} \rangle \rangle$ decay $\propto 1/V^\zeta$ with $\zeta \approx 1$ for the average, and $0.6 \lesssim \zeta < 1.0$ for the maximal difference. In contrast, the eigenstate-to-eigenstate fluctuations in the absence of single-particle quantum chaos are plotted in Fig.~\ref{figS3}(c) and~\ref{figS3}(d). It is apparent that even though the averages scale as in the regime with single-particle quantum chaos, the maximal outliers do not decrease with $V$. The latter is a violation of the single-particle eigenstate thermalization.

\end{document}